\documentclass[
superscriptaddress,
nofootinbib,
amsmath,amssymb,
aps,
twocolumn,
floatfix,
showpacs,
showkeys,
]{revtex4-2}

\usepackage[pdftex]{graphicx}
\usepackage{dcolumn}
\usepackage{orcidlink}
\usepackage{bm}
\usepackage{mathrsfs}
\usepackage{xtab,afterpage}
\usepackage{longtable}
\usepackage{color}
\usepackage{hyperref}
\usepackage{tikz}
\usepackage{comment}
\usepackage{multirow}
\usepackage{physics}
\usepackage{ulem}
\usepackage{tablefootnote}

\hypersetup{colorlinks,breaklinks,linkcolor=blue,urlcolor=blue,anchorcolor=blue,citecolor=blue}

\newcolumntype{d}[1]{D{.}{.}{#1}}

\makeatletter
\def\LT@LR@e{\LTleft\z@   \LTright\z@}
\LTcapwidth=\textwidth
\makeatother

\begin{document}

\title{Triaxiality and Shape Dynamics in $^{70}$Ge}

\author{T. M. Kowalewski\,\orcidlink{0000-0001-7493-1842}}
\email{tyler.kowalewski@unc.edu}
\author{A. D. Ayangeakaa\,\orcidlink{0000-0003-1679-3175}}
\email{ayangeak@unc.edu}
\affiliation{Department of Physics and Astronomy, University of North Carolina Chapel Hill, NC 27599, USA}
\affiliation{Triangle Universities Nuclear Laboratory, Duke University, Durham, NC 27708, USA}

\author{N. Sensharma\,\orcidlink{0000-0002-5046-9451}}\thanks{Present address: Physics Division, Argonne National Laboratory, Argonne, IL 60439, USA}
\author{R. V. F. Janssens\,\orcidlink{0000-0002-1874-7108}}
\affiliation{Department of Physics and Astronomy, University of North Carolina Chapel Hill, NC 27599, USA}
\affiliation{Triangle Universities Nuclear Laboratory, Duke University, Durham, NC 27708, USA}

\author{Y. M. Wang\,\orcidlink{0009-0001-8726-6324}}
\author{Q. B. Chen\,\orcidlink{0000-0001-5159-4468}}
\affiliation{Department of Physics, East China Normal University, Shanghai 200241, China}

\author{J. M. Allmond\,\orcidlink{0000-0001-6533-8721
}}
\affiliation{Physics Division, Oak Ridge National Laboratory, TN 37831-6371, USA}

\author{C. M. Campbell\,\orcidlink{0000-0003-3577-4316}}
\affiliation{Nuclear Science Division, Lawrence Berkeley National Laboratory, Berkeley, CA 94720, USA} 

\author{S. Carmichael\,\orcidlink{0000-0002-7454-1876}}
\affiliation{Department of Physics and Astronomy, University of Notre Dame, Notre Dame, IN 46556, USA}

\author{M. P. Carpenter\,\orcidlink{0000-0002-3237-5734}}
\author{P. Copp\,\orcidlink{0000-0002-4786-2404}}\thanks{Present address: Los Alamos National Laboratory, Los Alamos, New Mexico 87545, USA}
\affiliation{Physics Division, Argonne National Laboratory, Lemont, IL 60439, USA}

\author{C. Cousins\,\orcidlink{0009-0008-3855-8217}}
\affiliation{Department of Physics, University of Surrey, Surrey GU2 7XH, United Kingdom}

\author{M. Devlin\,\orcidlink{0000-0002-6948-2154}}
\affiliation{Physics Division, Los Alamos National Laboratory, Los Alamos, New Mexico 87545, USA}

\author{U. Garg\,\orcidlink{0000-0001-8751-4204}}
\affiliation{Department of Physics and Astronomy, University of Notre Dame, Notre Dame, IN 46556, USA}

\author{C. M{\"u}ller-Gatermann\,\orcidlink{0000-0002-9181-5568}}
\affiliation{Physics Division, Argonne National Laboratory, Lemont, IL 60439, USA}

\author{T. J. Gray\,\orcidlink{0000-0003-3965-6130}}\thanks{Present address: Department of Physics and Astronomy, University of Tennessee, Knoxville, TN 37966, USA}
\affiliation{Physics Division, Oak Ridge National Laboratory, TN 37831-6371, USA}

\author{D. J. Hartley\,\orcidlink{0000-0001-6295-6815}}
\affiliation{Department of Physics, United States Naval Academy, Annapolis, MD 21402, USA}

\author{J. Heery\,\orcidlink{0000-0002-3023-9907}}
\affiliation{Department of Physics, University of Surrey, Surrey GU2 7XH, United Kingdom}

\author{J. Henderson\,\orcidlink{0000-0002-6010-9644}}
\affiliation{Department of Physics, University of Surrey, Surrey GU2 7XH, United Kingdom}

\author{H. Jayatissa\,\orcidlink{0000-0001-8746-0234}}\thanks{Present address: Los Alamos National Laboratory, Los Alamos, New Mexico 87545, USA}
\affiliation{Physics Division, Argonne National Laboratory, Lemont, IL 60439, USA}

\author{S. R. Johnson\,\orcidlink{0009-0004-3440-5070}}
\affiliation{Department of Physics and Astronomy, University of North Carolina Chapel Hill, NC 27599, USA}
\affiliation{Triangle Universities Nuclear Laboratory, Duke University, Durham, NC 27708, USA}

\author{S. P. Kisyov\,\orcidlink{0000-0002-4718-7744}}\thanks{Present address: Lawrence Berkeley National Laboratory, Berkeley, CA 94720, USA}
\affiliation{Physics Division, Lawrence Livermore National Laboratory, Livermore, California 94550, USA}

\author{F. G. Kondev\,\orcidlink{0000-0002-9567-5785}}
\author{T. Lauritsen\,\orcidlink{0000-0002-9560-0388}}
\author{S. Nandi\,\orcidlink{0000-0003-0574-8467}}
\affiliation{Physics Division, Argonne National Laboratory, Lemont, IL 60439, USA}

\author{R. Rathod}
\affiliation{Department of Physics and Astronomy, University of Notre Dame, Notre Dame, IN 46556, USA}

\author{W. Reviol\,\orcidlink{0000-0002-5372-7743}}
\affiliation{Physics Division, Argonne National Laboratory, Lemont, IL 60439, USA}

\author{M. Rocchini\,\orcidlink{0000-0001-6869-0181}}\thanks{Present address: INFN Sezione di Firenze, Firenze I-50019, Italy}
\affiliation{Department of Physics, University of Guelph, Guelph, Ontario N1G 2W1, Canada}

\author{E. Rubino\,\orcidlink{0000-0002-6422-7015}}
\affiliation{Facility for Rare Isotope Beams, Michigan State University, East Lansing, Michigan 48824, USA}

\author{R. Russell\,\orcidlink{0000-0001-6580-0111}}
\affiliation{Department of Physics, University of Surrey, Surrey GU2 7XH, United Kingdom}

\author{A. Saracino\,\orcidlink{0009-0002-0357-2429}}
\affiliation{Department of Physics and Astronomy, University of North Carolina Chapel Hill, NC 27599, USA}
\affiliation{Triangle Universities Nuclear Laboratory, Duke University, Durham, NC 27708, USA}

\author{D. Seweryniak\,\orcidlink{0009-0003-3028-9693}}
\author{M. Siciliano\,\orcidlink{0000-0002-4598-0298}}
\affiliation{Physics Division, Argonne National Laboratory, Lemont, IL 60439, USA}

\author{C. Y. Wu\,\orcidlink{0000-0001-5772-7196}}
\affiliation{Physics Division, Lawrence Livermore National Laboratory, Livermore, California 94550, USA}

\date{\today}

\begin{abstract}
The electromagnetic properties of low-lying states in $^{70}$Ge were investigated via multi-step Coulomb excitation of a $^{70}$Ge beam impinging on a $^{208}$Pb target at the ATLAS facility of the Argonne National Laboratory. A total of 27 transitional elements and six diagonal matrix elements coupling 11 low-lying states, were extracted from the measured cross sections. These were used to calculate reduced transition probabilities, spectroscopic quadrupole moments, and rotational invariant shape parameters, providing enhanced precision and expanding on previous studies. The experimental data were compared within several theoretical frameworks, including the generalized triaxial rotor model, configuration interaction shell-model calculations, and computations within the combined frameworks of relativistic density functional theory and the five-dimensional collective Hamiltonian. The results demonstrate a good agreement with the experimental data and, in conjunction with calculations using a two-state mixing model, support significant triaxiality and strong mixing between the $0^+_1$ and $0^+_2$ states. This results in the magnitudes of their respective quadrupole deformations $[\beta_\text{rms}(0^+_1) = 0.228\,(3),\,\beta_\text{rms}(0^+_2) = 0.273\,(1)]$ being more similar than previously observed. The implications of these results for understanding the complex shape coexistence phenomena, the role of triaxiality, and shape evolution along the Ge isotopic chain are discussed.
\end{abstract}

\maketitle

\section{Introduction} \label{sec:intro} 

    The low-lying structure of the even-even Ge isotopes exhibits unique phenomena which have recently become amenable to tests with advanced theoretical models, making them the subject of numerous contemporary experimental investigations. The most striking features in these isotopes are evidence for shape coexistence, characterized by the presence of low-lying $0^+$ levels with properties markedly different from their respective ground states; the absence of a subshell closure at $N=40$, as evidenced by the highest energy $2^+_1$ state appearing in $^{70}$Ge rather than in $^{72}$Ge; and an apparent shape transition across the isotopic chain featuring weakly and strongly deformed structures associated with varying degrees of axial asymmetry \cite{Garrett2021}. This rich variety of phenomena within a single isotopic chain has made the even-even Ge isotopes important for testing and challenging current models. 
    
    During the past decade, targeted experiments that comprehensively investigated even-even Ge isotopes have been conducted, with the specific aim of precisely elucidating the interplay of shell structure evolution and collective behavior as a function of isospin near the $N=40$ subshell closure. These studies began with a set of Coulomb excitation experiments designed to probe the quadrupole collectivity of the low-lying states in $^{72}$Ge \cite{Ayangeakaa2016,Kotlinski1990} and explore the nature of triaxial deformation in $^{76}$Ge \cite{Ayangeakaa2019,Ayangeakaa2023}. In the former, evidence for the coexistence of two distinct, triaxially-deformed configurations associated with the $0^+_1$ and $0^+_2$ states was found. For $^{76}$Ge, the comprehensive set of $E2$ matrix elements extracted from the measured differential Coulomb excitation cross sections was compared to results of configuration-interaction shell-model calculations and computations carried out within the framework of the generalized triaxial rotor model (GTRM). The results demonstrated a remarkable agreement with the experimental data and, in conjunction with rotational invariant shape parameters derived via the Kumar-Cline sum rules \cite{KK1,KK2,KK3}, support a nearly maximum rigid triaxial deformation close to the ground state.
    
    Theoretically, the Ge isotopes have been extensively investigated using various computational approaches including the Hartree-Fock-Bogoliubov framework \cite{Parikh1972,Gaudefroy2009}, the Interacting Boson Model \cite{Hsieh1992, Nomura2017}, and covariant density functional theory \cite{Guo2007,Mennana2021}. These studies generally support the interpretation that multiparticle excitations across the $Z=28$ proton shell gap, combined with intrusion of the $g_{9/2}$ neutron orbital into the lower subshells, contribute to the emergence of deformed configurations at low excitation energy. However, despite this progress, theoretical studies continue to face challenges in accurately accounting for some of the observations, such as the parabolic trend observed in the energy of the $0^+_2$ states across the stable Ge isotopes.
    
    Moreover, the precise evolution of deformation along the isotopic chain remains a topic of ongoing debate. Two competing shape configurations appear to coexist: a deformed, near-prolate (triaxial) one, which is the $0^+_1$ ground state in $^{76}$Ge, and another near-spherical (triaxial) one associated with the $0^+_2$ level. These two configurations are proposed to switch roles with decreasing neutron excess. Analyses of pickup and stripping reactions suggest that this transition occurs between $^{72}$Ge and $^{74}$Ge \cite{Carchidi1984}, while Coulomb excitation measurements from 1980 \cite{Lecomte1980} point to it occurring between $^{70}$Ge and $^{72}$Ge. Furthermore, a more recent Coulomb excitation study of $^{70}$Ge performed by Sugawara \textit{et al.}~\cite{Sugawara2003} reported an unexpectedly large deformation for the $0^+_2$ state, with a rotational invariant $\expval{Q^2}$  of $0.64(9)$ e$^2$b$^2$ (See caption of Fig. 35 in Ref.~\cite{Heyde2011}). This corresponds to a quadrupole deformation parameter $\beta_\mathrm{rms} = 0.43(3)$, marking a significant deformation at low excitation energy and spin. While such significant quadrupole deformations have been observed in nuclei near the $N = 40$ subshell closure, they are predominantly seen at high spin and are typically associated with highly-deformed or superdeformed rotational bands. Such sequences have yet to be observed in $^{70}$Ge. On the other hand, recent $^{74}$Ge Coulomb excitation data~\cite{nirupama2024} indicate that the deformations associated with the $0^+_{1}$ and $0^+_{2}$ states are much more similar than previously reported \cite{Toh2013}. These results raise questions about the detailed interplay between these competing configurations. A precise characterization of the evolution of quadrupole deformation and axial symmetry, along with a better understanding of the roles played by competing shape configurations in the even-even Ge isotopes, is, therefore, desirable to unravel the complex collective phenomena that govern shape dynamics in this region of the nuclear chart.
    
    In the present investigation, a comprehensive study of quadrupole collectivity associated with the low-lying states in $^{70}$Ge is presented. The deformation parameters, extracted from the $E2$ matrix elements obtained from a multi-step Coulomb excitation measurement, are compared with the results of several theoretical models, including the generalized triaxial rotor model (GTRM), the configuration interaction shell model, and computations performed within the combined frameworks of the relativistic density functional theory and the five-dimensional collective Hamiltonian (5DCH). In addition, an analysis using a two-state mixing model suggests significant mixing between the configurations associated with the $0^+_1$ and $0^+_2$ states. The implications of these results for understanding the nature of shape coexistence phenomena and of shape evolution along the Ge isotopic chain are discussed.
    
    The present article is organized as follows. First, details of the experimental setup and methodology are presented in Section \ref{sec:expt}. Next, the analysis and the results, including the observed spectra, level scheme, and final set of matrix elements are given in Sec.~\ref{sec:analysis}. A discussion of the results and comparisons with collective rotor models, large-scale shell-model calculations in the jj44 model space, a two-state mixing model, and calculations performed in the framework of relativistic density functional theory, are covered in Section \ref{sec:discussion}. Finally, Sec.~\ref{sec:conclusion} provides a summary and final conclusions.

\section{Experimental Details} \label{sec:expt} 

\begin{figure}[!t]
        \centering
        \includegraphics[width=\columnwidth,trim={0.5cm 0.5cm 0.5cm 0.5cm},clip]{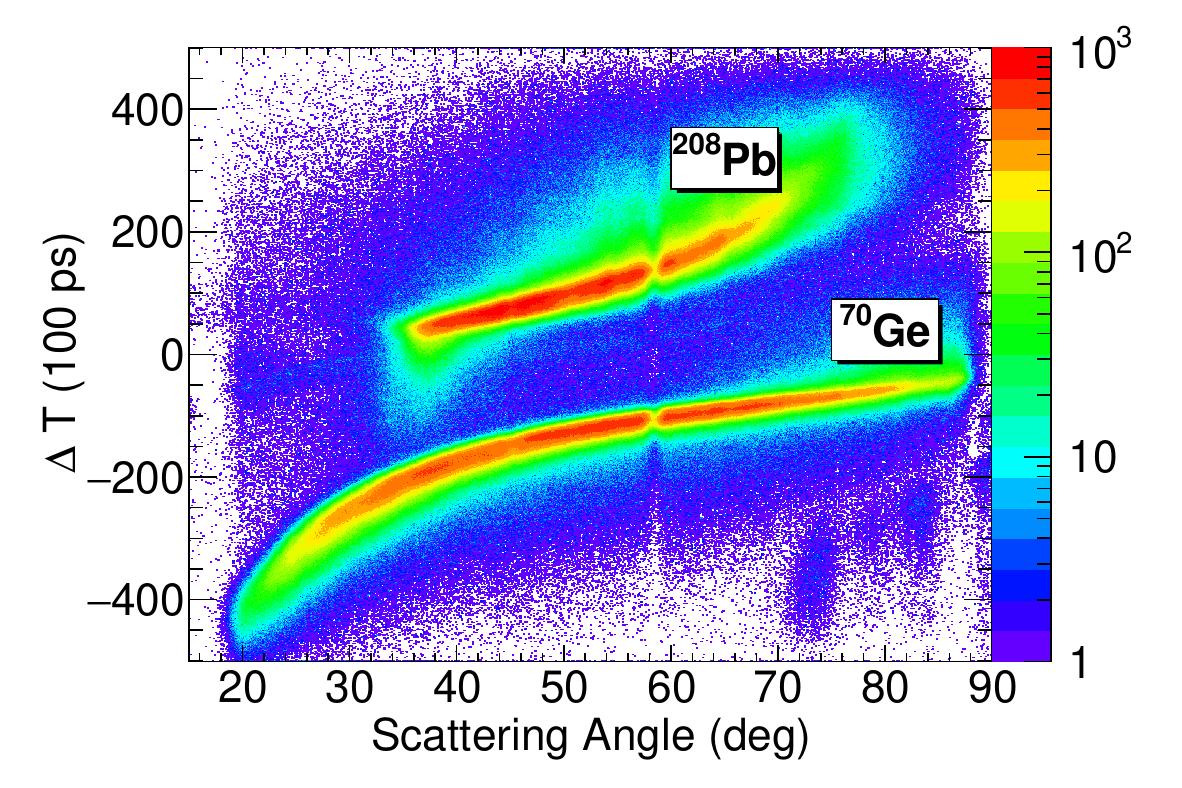} 
        \caption{(Color online) Difference in the time-of-flight between the scattered reaction participants as a function of the laboratory scattering angle measured with the CHICO2 detector, illustrating the clear separation between the detected $^{208}$Pb and $^{70}$Ge ions. The color scale (right side) provides the range of z-axis values (counts/$1.55^\circ$) in this plot. The CHICO2 detector system is composed of left and right hemispheres. The plot above shows the time-of-flight difference $\Delta T = t_\mathrm{left} - t_\mathrm{right}$ between the hemispheres for sequential hits at forward scattering angles. A similar pattern and separation is achieved in the plot for the other hemisphere, while, due to the scattering kinematics, $^{208}$Pb is absent from the backward angles in both hemispheres.}
        \label{fig:partspec}
    \end{figure}

    Excited states in $^{70}$Ge were populated via multi-step Coulomb excitation in an experiment performed at the Argonne Tandem Linac Accelerator System (ATLAS) accelerator facility at Argonne National Laboratory (ANL). A beam of $^{70}$Ge ions at an energy of $297$ MeV bombarded a $0.442$-mg/cm$^2$ thick, isotopically-enriched $^{208}$Pb target sandwiched between a 6-$\mu$g/cm$^2$ $^\text{nat}$Al front layer and a 40-$\mu$g/cm$^2$ $^\text{nat}$C backing. This beam energy was chosen just above the Cline safe energy \cite{Cline1969} to ensure a purely electromagnetic interaction for most scattering angles\footnote{This beam energy corresponds to laboratory scattering angles up to $\approx 140^\circ$ being considered safe. At the maximum scattering angle considered ($160^\circ$), the beam energy is only $\approx 2.5\,\%$ above the corresponding safe energy.}, while also enabling increased population of higher spin/energy states. The $\gamma$ rays emitted during de-excitation were detected using the GRETINA array \cite{Paschalis2013} in kinematic coincidence with CHICO2 \cite{Simon2000, CHICO2}, a large solid angle, position-sensitive, parallel-plate avalanche counter. CHICO2 is capable of maintaining a mass resolution $(\Delta m/m)$ of $\approx 5\,\%$ and covers laboratory scattering angles between $20^\circ$ and $88^\circ$ downstream from the target and $96^\circ$ and $164^\circ$ upstream, with $1.55^\circ$ and $2.46^\circ$ localization in $\theta$ (polar angle) and $\phi$ (azimuthal angle), respectively. In addition, the system achieves a timing resolution of $\approx1.2$ ns (FWHM), which is sufficient to measure time-of-flight differences ($\Delta T$) between the reaction participants as a function of the laboratory polar scattering angle, which is critical for identifying reaction participants at forward angles. A two-dimensional histogram plotting $\Delta T$ against the laboratory scattering angle provides a clear separation between the reaction products, as seen in Fig.~\ref{fig:partspec}. This allows for clean gates to be placed on events coincident with either the scattered beam or recoiling target particles. In the present work, gates were drawn on scattered $^{70}$Ge ions only.
    
\begin{figure*}[!t] 
        \centering
      \includegraphics[width=\textwidth]{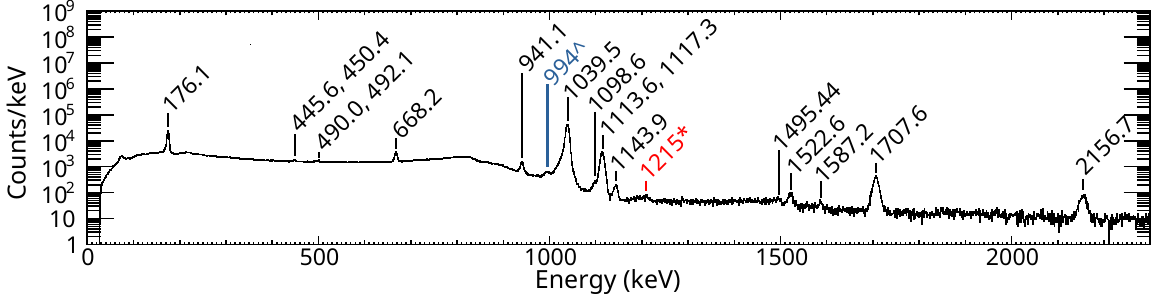} 
        \caption{(Color online) Doppler- and efficiency-corrected $\gamma$-ray spectrum measured with GRETINA following the Coulomb excitation of a $^{70}$Ge beam incident on a $^{208}$Pb target. This spectrum shows $\gamma$ rays that were coincident with $^{70}$Ge recoils detected by CHICO2 at laboratory scattering angles between $130^\circ$ and $160^\circ$; it spectrum contains all transitions observed in the experiment. The typical energy resolution obtained ranges from a full width at half maximum (FWHM) of $\approx 3$ keV at 668 keV to $\approx 7$ keV at 2157 keV. The $994$-keV peak (shown in blue and marked with a caret) was excluded from the analysis, as it was not observed in previous works and its placement in the level scheme is uncertain. The $1215$-keV peak (shown in red and marked with an asterisk) was also excluded, as it is believed to correspond to the sum peak resulting from simultaneous detection of $1039.5$-keV ($2^+_1\to0^+_1$) and $176.1$-keV ($0^+_2\to2^+_1$) $\gamma$ rays.}
        \label{fig:repspec}
    \end{figure*}

    The time and position resolutions of the CHICO2 detector, in combination with $\gamma$-ray position information from GRETINA, enabled event-by-event reconstruction of the two-body reaction kinematics and precise Doppler correction of the $\gamma$ rays emitted in flight. The Doppler correction followed the standard procedure routinely employed in GRETINA-CHICO2 experiments \cite{Henderson2025,Little2022,Bucher2016}. In this approach, CHICO2 provides the trajectories of the reaction products, allowing the determination of the angle between the detected $\gamma$ ray and the scattered $^{70}$Ge ion on an event-by-event basis \cite{CHICO2}.While GRETINA is capable of $\gamma$-ray tracking, which allows event-by-event reconstruction of the Compton scattered $\gamma$ rays, only the non-tracked spectra were used in this study. Consequently, each signal from GRETINA was treated as an independent $\gamma$-ray hit. Fig.~\ref{fig:repspec} presents the Doppler-reconstructed spectrum of all $\gamma$ rays measured by GRETINA in coincidence with scattered $^{70}$Ge ions detected by CHICO2.
    
\begin{figure}[!t]
        \centering  \includegraphics[width=\columnwidth, trim={5.5cm 5.1cm 5.2cm 4.4cm}, clip]{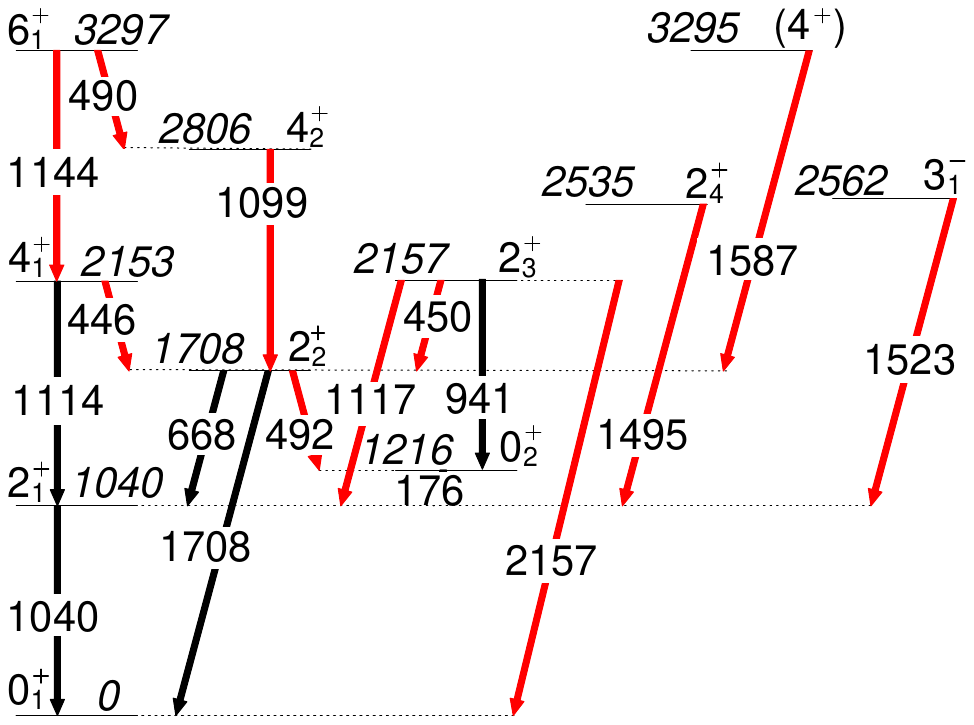}
        \caption{(Color online) Partial level scheme of $^{70}$Ge including all the transitions and levels observed in the present study. Transitions in black are those observed in Ref. \cite{Sugawara2003} while those in red were known in the literature, but observed in Coulomb excitation for the first time in the present work. Spin, parity, and energy assignments for the observed levels were adopted from the most recent ENSDF evaluation \cite{Mukhopadhyay2017}, with tentative values enclosed in parentheses and energies rounded to the nearest keV. Note that the 3295-keV state is listed as $3^+,4^+$ in the latest ENSDF evaluation, but taken to be $4^+$ in the present analysis as Coulomb excitation preferentially excites through $E2$ transitions and GOSIA requires a definite spin assignment.}
        \label{fig:levelscheme}
    \end{figure}

    A total of seventeen (17) unique $\gamma$ rays connecting eleven (11) low-lying excited states in $^{70}$Ge were observed. An additional $\gamma$ ray at $994$ keV was observed, but not included in the analysis, since it had not been previously reported and its placement in the level scheme is uncertain. The peak at $1215$ keV was also excluded from the analysis as it is believed to be the sum peak resulting from simultaneous detection of the $1039.5$-keV ($2^+_1\to0^+_1$) and $176.1$-keV ($0^+_2\to2^+_1$) $\gamma$ rays. A partial level scheme incorporating all the observed transitions is given in Fig.~\ref{fig:levelscheme}. Transitions marked in black are those observed by Sugawara \textit{et al.} \cite{Sugawara2003} in the most recent $^{70}$Ge Coulomb excitation measurement prior to this work. The transitions in red are known from the literature \cite{Gurdal2016}, but had not been seen in Coulomb excitation prior to this work. Spin and parity assignments of the observed levels are adopted from the ENSDF database \cite{Gurdal2016}.

    \begin{figure*}[htbp]
        \centering
        \includegraphics[width=\textwidth]{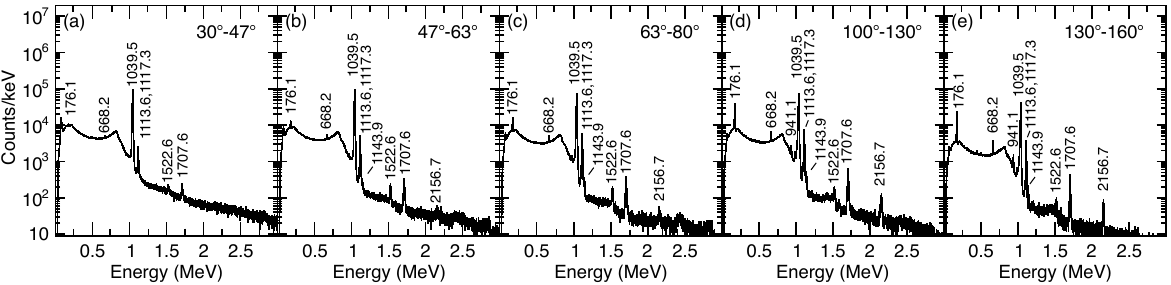} 
        \caption{Doppler-corrected $\gamma$-ray decay spectra produced from the Coulomb excitation of $^{70}$Ge, shown in five subsets corresponding to different projectile laboratory scattering angles. At backward angles more $\gamma$ rays are seen, particularly from transitions including higher energy and spin states. This qualitatively shows the dependence of the excitation cross section on scattering angle, energy, and spin.}
        \label{fig:multispec}
    \end{figure*}

\section{Data Analysis and Results}\label{sec:analysis}

    Reduced transition matrix elements linking the nuclear states govern the excitation and decay processes involved in multi-step Coulomb excitation. In addition, they provide insight into the deformation of the nucleus of interest.  Through the minimization of a standard $\chi^2$ function of the measured and calculated $\gamma$-ray yields, branching ratios, and lifetimes, a set of ($E1$, $E2$, $E3$, and $M1$) matrix elements was obtained. The calculated $\gamma$-ray yields were obtained with the semi-classical, multi-step Coulomb excitation code GOSIA \cite{Czosnyka1983,GosiaManual2012} and were normalized to that of the $2^+_1\to0^+_1$ transition. Minimization was performed with the GOSIAFitter program \cite{GOSIAFitter}. Additionally, a set of spectroscopic data taken from the most recent ENSDF evaluation \cite{Mukhopadhyay2017} were also used to complement the measured quantities and further constrain the fit (see Table \ref{tab:spectroscopicdata}). Since the excitation probability depends on the relative phases, signs, and magnitudes of the $E2$ matrix elements, multiple initial sets of matrix elements, which sampled all possible signs of the quadrupole interference term \cite{Allmond2009}, were examined. This approach further increases the likelihood that the set of matrix elements found is representative of the global minimum of the $\chi^2$ surface. In addition, the $\gamma$-ray yields were corrected for possible internal conversions, the finite size and relative efficiency of the $\gamma$-ray detectors, and the attenuation of particle-$\gamma$ correlations due to deorientation effects during recoil in vacuum. The BrIcc database \cite{Kibedi2008} was used to compute internal conversion coefficients for all observed transitions. The $0^+_2\to0^+_1$ $E0$ transition was not considered in the GOSIA analysis, as the latest ENSDF evaluation \cite{Gurdal2016} lists an $E0/E2$ branching ratio of $1\,\%$, which is well below the existing systematic uncertainty introduced by approximations made by GOSIA in the yield calculations.

    \begin{table}[!b]
        \caption{Lifetimes $(\tau)$ and branching ratios used to constrain the minimization process. All values were taken from the most recent ENSDF evaluation \cite{Mukhopadhyay2017}.\footnote{The lifetime values chosen to constrain the fit are those listed in the evaluation as having been measured via Coulomb excitation. Therefore, not all lifetimes in Table \ref{tab:spectroscopicdata} are the primary adopted value for a given state. These were chosen to limit the effect of systematic differences between Coulomb excitation and other methodologies.} The uncertainties are symmetrized for use in GOSIA.}
        \label{tab:spectroscopicdata}
        \begin{ruledtabular}
        \begin{tabular}{ccccc}
            $I^\pi$ & $\tau$ (ps)  & & $I_i^\pi \to I_{f1}^\pi/I^\pi_i\to I^\pi_{f_0}$ & Branching Ratio \\ \hline 
            $2^+_1$	& 1.89 (3)      & & $2^+_2\to0^+_1/2^+_2\to2^+_1$ & 0.792 (53)  \\
            $4^+_1$	& 2.45 (58)     & & $2^+_2\to0^+_2/2^+_2\to2^+_1$ & 0.049 (5) \\
            $2^+_2$	& 6.1 (38)      & & $4^+_1\to2^+_2/4^+_1\to2^+_1$ & 0.007 (4) \\
            $0^+_2$	& 6925 (1010)   & & $2^+_3\to0^+_2/2^+_3\to2^+_1$ & 0.62 (5)  \\
       		        &              & & $2^+_3\to0^+_1/2^+_3\to2^+_1$ & 0.171 (14) \\
    		        &              & & $2^+_3\to2^+_2/2^+_3\to2^+_1$ & 0.047 (24) \\
                    &              & & $4^+_2\to4^+_1/4^+_2\to2^+_2$&0.118 (14)\\
        \end{tabular}  
       \end{ruledtabular}
    \end{table}
    
    To enhance the sensitivity to the matrix elements and exploit the dependence of the excitation probability on the projectile's scattering angle, the data were partitioned into five angular subsets corresponding to the laboratory scattering angles $30^\circ - 47^\circ$, $47^\circ-63^\circ$, $63^\circ-80^\circ$, $100^\circ-130^\circ$, and $130^\circ-160^\circ$. Gating on $\gamma$ rays which were coincident with scattered $^{70}$Ge detections (see Fig. \ref{fig:partspec}) resulted in a total of fifty-two (52) efficiency-corrected, $\gamma$-ray yields. Sample spectra, which demonstrate the angular dependence, are found in Fig \ref{fig:multispec}. As the average scattering angle increases from $\approx 39^\circ$ [Fig.~\ref{fig:multispec}(a)] to $\approx 145^\circ$ [Fig.~\ref{fig:multispec}(e)], $\gamma$ rays associated with the de-excitation from higher-lying states become apparent.
    
\begin{figure}[htbp!]
        \includegraphics[width=\columnwidth]{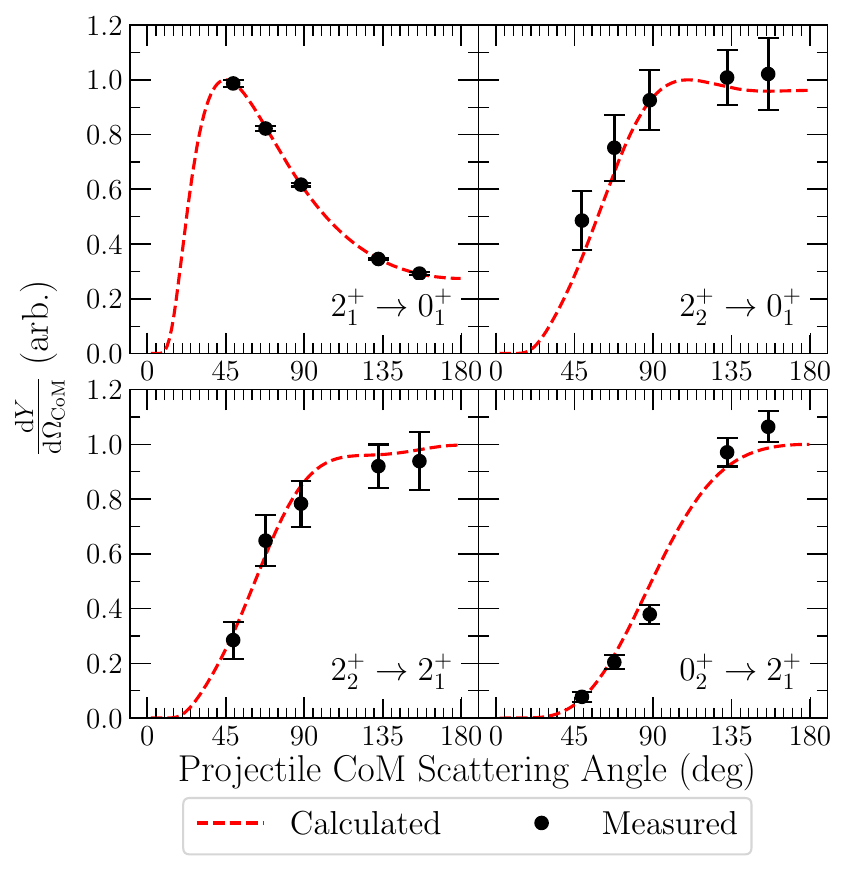} 
        \caption{(Color online) Comparison between the measured differential $\gamma$-ray yields (black points) and the differential yields calculated using the set of matrix elements obtained from the GOSIA fit (red dashed curve) for several representative transitions. Each point corresponds to the center of the five aforementioned angle partitions, converted into the center of mass (CoM) frame. In this frame, the characteristic shape of the differential yields -- which depends on energy, spin, and parity of the initial and final levels -- can be more distinctly seen. }
        \label{fig:yields_comparison}
    \end{figure}

    The final set of transition and diagonal matrix elements, which best reproduce the experimental $\gamma$-ray yields and known spectroscopic information, is displayed in Table \ref{tab:results}. This table includes all couplings included in the GOSIA analysis, excluding the fixed couplings to ``buffer" states, which were set to values derived from the states' lifetimes, where available.\footnote{A single ``buffer" state was added to the top of each band in the GOSIA coupling scheme to ensure that contributions to the population of the highest-lying observed states in each band, due to the decay or virtual excitation of higher-lying states, were accurately calculated. For more information view Section 6.8 of Ref. \cite{GosiaManual2012}.} A representative sample of the yields calculated from the set of fitted matrix elements is compared to the measured yields in Fig. \ref{fig:yields_comparison}. Convergence was achieved with a reduced $\chi^2$ of $0.906$. A total of thirty-three (33) reduced matrix elements including one (1) $E1$, twenty-six (26) $E2$, one (1) $E3$, and five (5) $M1$ ones were determined. For all listed matrix elements, the quoted uncertainties were derived in the standard way using GOSIA: a probability distribution was constructed in the space of fitted parameters and the total probability was requested to be equal to the chosen confidence limit (in this case $1\sigma$) while contributions to the total uncertainty which arise from cross-correlation effects were also taken into account \cite{GosiaManual2012}. These uncertainties include statistical, as well as systematic ones, originating from the nature of the detector system. An additional $5\,\%$ contribution was added to the quoted uncertainties to account for systematic error introduced by approximations used by the GOSIA code during $\gamma$-ray yield calculations \cite{GosiaManual2012}.
    
    \begin{longtable*}{@{\extracolsep{\fill}} cccccccc @{\extracolsep{\fill}}}
    
        \caption{Summary of all $E1$, $E2$, $E3$, and $M1$ matrix elements and reduced transition probabilities for $^{70}$Ge deduced in the present work. Units for the matrix elements are eb$^{1/2}$, eb, eb$^{3/2}$ and $\mu_N$ for $E1$, $E2$, $E3$, and $M1$ transitions, respectively. Accordingly, $M1$, $E1$, $E2$, and $E3$ reduced transition probabilities are listed in units of $\mu^2_N$, e$^2$b, e$^2$b$^2$, and e$^2$b$^3$, respectively. Here, $\lambda$ stands for electric ($E$) or magnetic ($M$) and the multipole order $L$ takes the values of either $1$, $2$, or $3$. The initial and final levels for a transition are denoted by the same $I^\pi_n$ labels as in Fig. \ref{fig:levelscheme}. Spectroscopic quadrupole moments $Q_s(I^\pi)$ are also included, where applicable, and are given in units of eb. The last two columns present the reduced transition probabilities in Weisskopf units (W.u.). Upper limits are given at the $1\sigma$ confidence level. For reference, the ENSDF-adopted $\gamma$-ray energies (rounded to the nearest tenth of a keV) are also listed with their respective transitions, where available. Note that the uncertainties are quoted relative to the most insignificant digit of the value and in a format based on whether the errors are symmetric.
        \label{tab:results}}\\
        
        \hline \hline
        \multicolumn{1}{c}{\multirow{2}{*}{$I^\pi_i \rightarrow I^\pi_f $}} & \multicolumn{1}{c}{\multirow{2}{*}{$E_\gamma$ [keV]}} & \multicolumn{1}{c}{\multirow{2}{*}{Mult.}} & \multicolumn{2}{c}{$\left < I_i^\pi || M(\lambda L)|| I_f^\pi \right >\uparrow$} & \multicolumn{1}{c}{$B(\lambda L\downarrow;\, I_f^\pi \rightarrow I_i^\pi )$} & \multicolumn{2}{c}{$B(\lambda L\downarrow;\, I_f^\pi \rightarrow I_i^\pi)$ [W.u.]} \\ \cline{4-5} \cline{7-8} 
        \multicolumn{1}{c}{} & \multicolumn{1}{c}{} & \multicolumn{1}{c}{} & \multicolumn{1}{c}{This work} & \multicolumn{1}{c}{Sugawara {\it et al.}~\cite{Sugawara2003}} & \multicolumn{1}{c}{or $Q_s(I^\pi)$} & \multicolumn{1}{c}{This Work} & \multicolumn{1}{c}{Sugawara {\it et al.}~\cite{Sugawara2003}}\\
        \hline
    
        \endfirsthead
    
        \hline \hline
        \multicolumn{1}{c}{\multirow{2}{*}{$I^\pi_i \rightarrow I^\pi_f $}} & \multicolumn{1}{c}{\multirow{2}{*}{$E_\gamma$ [keV]}} & \multicolumn{1}{c}{\multirow{2}{*}{Mult.}} & \multicolumn{2}{c}{$\left < I_i^\pi || M(\lambda L)|| I_f^\pi \right >\uparrow$} & \multicolumn{1}{c}{$B(\lambda L\downarrow;\, I_f^\pi \rightarrow I_i^\pi )$} & \multicolumn{2}{c}{$B(\lambda L\downarrow;\, I_f^\pi \rightarrow I_i^\pi)$ [W.u.]} \\ \cline{4-5} \cline{7-8} 
        \multicolumn{1}{c}{} & \multicolumn{1}{c}{} & \multicolumn{1}{c}{} & \multicolumn{1}{c}{This work} & \multicolumn{1}{c}{Sugawara {\it et al.}~\cite{Sugawara2003}} & \multicolumn{1}{c}{or $Q_s(I^\pi)$} & \multicolumn{1}{c}{This Work} & \multicolumn{1}{c}{Sugawara {\it et al.}~\cite{Sugawara2003}}\\
        \hline
    
        \endhead

        \\
        \hline\hline
        \endfoot

        \\
        \hline \hline
        \endlastfoot
    
        \rule{0pt}{3ex}
        $0^+_1$ $\to$ $2^+_1$ & 1039.5 & $E2$ & $0.422(4)$ & $0.426(5)$ & $0.0356(7)$ & $20.8(4)$ & $21.2(5)$ \\
        \rule{0pt}{3ex}
        $0^+_1$ $\to$ $3^-_1$ &  &$E3$ & $0.23\left(_{-3}^{+2}\right)$ &  & $0.008\left(_{-2}^{+1}\right)$ & $26\left(_{-7}^{+5}\right)$ &  \\
        \rule{0pt}{3ex}
        $0^+_1$ $\to$ $2^+_2$ & 1707.6 &$E2$ & $-0.0526(8)$ & $-0.0434(13)$ & $0.00055(2)$ & $0.32(1)$ & $0.22(1)$ \\
        \rule{0pt}{3ex}
        $0^+_1$ $\to$ $2^+_3$ & 2156.7 &$E2$ & $0.0232(7)$ & $0.027(3)$ & $0.00011(1)$ & $0.063(4)$ & $0.09(2)$ \\
        \rule{0pt}{3ex}
        $0^+_1$ $\to$ $2^+_4$ &  &$E2$ & $-0.02\left(_{-1}^{+5}\right)$ &  & $<0.00005$ & $<0.3$ & \\
        \rule{0pt}{3ex}
        $2^+_1$ $\to$ $2^+_1$ &  &$E2$ & $0.22\left(_{-3}^{+4}\right)$ & $0.05(4)$\footnotemark[1] & $0.17\left(_{-2}^{+3}\right)$ &  &  \\
        \rule{0pt}{3ex}
        $2^+_1$ $\to$ $3^-_1$ & 1522.6 &$E1$ & $-0.044\left(_{-3}^{+16}\right)$ &  & $0.00028\left(_{-4}^{+20}\right)$ & $0.026\left(_{-3}^{+18}\right)$ &  \\
        \rule{0pt}{3ex}
        $2^+_1$ $\to$ $4^+_1$ & 1113.6 &$E2$ & $0.68\left(_{-1}^{+5}\right)$ & $0.54(10)$ & $0.051\left(_{-2}^{+8}\right)$ & $30.0\left(_{-9}^{+44}\right)$ & $19(7)$ \\
        \rule{0pt}{3ex}
        $2^+_1$ $\to$ $0^+_2$ & 176.1 &$E2$ & $0.232\left(_{-11}^{+4}\right)$ & $0.272(11)$ & $0.054\left(_{-5}^{+2}\right)$ & $31\left(_{-3}^{+1}\right)$ & $43(3)$ \\
        \rule{0pt}{3ex}
        $2^+_1$ $\to$ $2^+_2$ & 668.2 &$E2$ & $0.416\left(_{-9}^{+13}\right)$ & $0.42(7)$ & $0.0346\left(_{-15}^{+21}\right)$ & $20.2\left(_{-9}^{+13}\right)$ & $21(7)$ \\
        \rule{0pt}{3ex}
        $2^+_1$ $\to$ $2^+_2$ & 668.2 &$M1$ & $-0.250\left(_{-17}^{+5}\right)$ &  & $0.0125\left(_{-17}^{+5}\right)$ & $0.0070\left(_{-10}^{+3}\right)$ &  \\
        \rule{0pt}{3ex}
        $2^+_1$ $\to$ $2^+_3$ & 1117.3 &$E2$ & $0.276\left(_{-7}^{+5}\right)$ & $0.53(6)$ & $0.0152\left(_{-8}^{+6}\right)$ & $8.9\left(_{-5}^{+3}\right)$ & $33(7)$ \\
        \rule{0pt}{3ex}
        $2^+_1$ $\to$ $2^+_3$ & 1117.3 &$M1$ & $-0.08\left(_{-1}^{+16}\right)$ & & $0.0013\left(_{-3}^{+51}\right)$ & $0.0007\left(_{-2}^{+29}\right)$ & \\
        \rule{0pt}{3ex}
        $2^+_1$ $\to$ $2^+_4$ & 1495.4 &$E2$ & $0.03\left(_{-5}^{+4}\right)$ &  & $<0.0007$ & $<0.4$ &  \\
        \rule{0pt}{3ex}
        $2^+_1$ $\to$ $2^+_4$ & 1495.4 & $M1$ & $<|1.0|$ &  & $<0.19$ & $<0.11$ &  \\
        \rule{0pt}{3ex}
        $2^+_1$ $\to$ $4^+_3$ & 2255.2 &$E2$ & $<|0.13|$ &  & $<0.0019$ & $<1.1$ &  \\
        \rule{0pt}{3ex}
        $4^+_1$ $\to$ $4^+_1$ & &$E2$ & $0.14\left(_{-6}^{+5}\right)$ & $0.29(7)$\footnotemark[1] & $0.11\left(_{-5}^{+4}\right)$ &  &  \\
        \rule{0pt}{3ex}
        $4^+_1$ $\to$ $6^+_1$ & 1143.9 &$E2$ & $0.75\left(_{-8}^{+6}\right)$ &  & $0.043\left(_{-9}^{+7}\right)$ & $25\left(_{-5}^{+4}\right)$ &  \\
        \rule{0pt}{3ex}
        $4^+_1$ $\to$ $4^+_2$ & 653.2 &$E2$ & $0.38\left(_{-3}^{+5}\right)$ &  & $0.016\left(_{-3}^{+4}\right)$ & $9.3\left(_{-15}^{+25}\right)$ &  \\
        \rule{0pt}{3ex}
        $4^+_1$ $\to$ $4^+_2$ & 653.2 &$M1$ & $0.22\left(_{-3}^{+4}\right)$ &  & $0.005\left(_{-1}^{+2}\right)$ & $0.0030\left(_{-8}^{+11}\right)$ &  \\
        \rule{0pt}{3ex}
        $4^+_1$ $\to$ $2^+_3$ &  &$E2$ & $0.59(4)$ & $0.86(12)$ & $0.069(9)$ & $41.0(6)$ & $86(24)$ \\
        \rule{0pt}{3ex}
        $6^+_1$ $\to$ $6^+_1$ & &$E2$ & $0.1\left(_{-4}^{+2}\right)$ &  & $0.1\left(_{-3}^{+1}\right)$ &  &  \\
        \rule{0pt}{3ex}
        $0^+_2$ $\to$ $2^+_2$ & 492.1 &$E2$ & $0.291\left(_{-8}^{+10}\right)$ & $0.25(2)$ & $0.0169\left(_{-9}^{+12}\right)$ & $9.9\left(_{-5}^{+7}\right)$ & $7(1)$ \\
        \rule{0pt}{3ex}
        $0^+_2$ $\to$ $2^+_3$ & 941.1 &$E2$ & $-0.349\left(_{-9}^{+8}\right)$ & $-0.71(13)$ & $0.0244\left(_{-13}^{+11}\right)$ & $14.21\left(_{-73}^{+65}\right)$ & $59(22)$ \\
        \rule{0pt}{3ex}
        $2^+_2$ $\to$ $4^+_1$ & 445.6 &$E2$ & $-0.43\left(_{-3}^{+2}\right)$ & $-0.52(12)$ & $0.021\left(_{-3}^{+2}\right)$ & $12\left(_{-2}^{+1}\right)$ & $18(8)$ \\
        \rule{0pt}{3ex}
        $2^+_2$ $\to$ $2^+_2$ & &$E2$ & $-0.44\left(_{-3}^{+4}\right)$ & $-0.09(5)$\footnotemark[1] & $0.33\left(_{-2}^{+3}\right)$ &  &  \\
        \rule{0pt}{3ex}
        $2^+_2$ $\to$ $4^+_2$ & 1098.5 &$E2$ & $0.44\left(_{-2}^{+3}\right)$ &  & $0.022\left(_{-2}^{+3}\right)$ & $12\left(_{-1}^{+2}\right)$ &  \\
        \rule{0pt}{3ex}
        $2^+_2$ $\to$ $2^+_3$ & 450.4 &$E2$ & $0.41\left(_{-4}^{+3}\right)$ & $0.23(9)$ & $0.034\left(_{-7}^{+5}\right)$ & $19\left(_{-4}^{+3}\right)$ & $6(5)$ \\
        \rule{0pt}{3ex}
        $2^+_2$ $\to$ $2^+_3$ & 450.4 &$M1$ & $0.204\left(_{-7}^{+21}\right)$ &  & $0.0083\left(_{-6}^{+17}\right)$ & $0.0047\left(_{-3}^{+10}\right)$ &  \\
        \rule{0pt}{3ex}
        $2^+_2$ $\to$ $(4^+)$ & 1587.2 &$E2$ & $<|0.64|$ &  & $<0.046$ & $<27$ &  \\
        \rule{0pt}{3ex}
        $4^+_2$ $\to$ $6^+_1$ & 490 &$E2$ & $<|0.9|$ & & $<0.062$ & $<36$ &  \\
        \rule{0pt}{3ex}
        $4^+_2$ $\to$ $4^+_2$ &  &$E2$ & $0.7\left(_{-2}^{+3}\right)$ &  & $0.5\left(_{-1}^{+2}\right)$ &  &  \\
        \rule{0pt}{3ex}
        $2^+_3$ $\to$ $2^+_3$ &  &$E2$ & $0.49\left(_{-8}^{+27}\right)$ & $0.34(13)$\footnotemark[1] & $0.37\left(_{-6}^{+20}\right)$ & & \\
    \end{longtable*}
    \footnotesize{\noindent \footnotemark[1]Calculated using the published spectroscopic quadrupole moment.}\normalsize

\section{Discussion}\label{sec:discussion}
    
    \subsection{Matrix Elements and Transition Probabilities}\label{sec:matrixelements}
    
        The present study provides a comprehensive set of matrix elements which couple the low-lying states of $^{70}$Ge and, thus, considerably expands upon the previously published data. Where applicable, these matrix elements are largely consistent with those from previous Coulomb excitation experiments \cite{Sugawara2003,Lecomte1980}, but with improved precision in nearly every case (see Table \ref{tab:results}). Notable differences are found in the magnitudes of the $\left<0^+_1||E2||2^+_2\right>$ matrix element, the diagonal elements (static moments), and all the elements which couple to the $2^+_3$ state. Many of the diagonal elements were observed to be larger than previously reported. For example, the $2^+_1$ static moment of $0.22_{-0.03}^{+0.04}$ eb is in better agreement with the $0.12(8)$ eb value reported in Ref. \cite{Lecomte1980} than the $0.05(4)$ eb one of Ref. \cite{Sugawara2003}. Similarly, the values of the $2^+_2$ and $2^+_3$ static moments were found to be significantly higher than those reported earlier: $-0.44\left(_{-4}^{+3}\right)$ and $0.49\left(_{-8}^{+27}\right)$ eb vs. $-0.09(5)$ and $0.34(13)$ eb in Ref. \cite{Sugawara2003}, respectively. Finally, while elements coupling the $2^+_3$ level to lower-lying states were included in the analysis of Ref. \cite{Sugawara2003}, only the $\gamma$-ray yield of the $2^+_3\to0^+_2$ transition was measured and used as a constraint in the fitting procedure. As a result, the values of the other transitions were only directly constrained by literature branching ratios. The measurement of these yields in the present analysis further constrains the corresponding matrix elements, which may explain the discrepancy between the present data and that of Ref. \cite{Sugawara2003}.
        
        Despite these discrepancies, the overall picture of the deformation and collectivity inferred from the transition strengths and static moments is consistent with previous works \cite{Sugawara2003,Lecomte1980}. Specifically, the presence of enhanced transition probabilities ($>20$ W.u) in the ground-state ($K=0$) band which increase with spin, and the positive sign of the static quadrupole moments, supports the interpretation of an oblate-deformed $0^+_1$ state (when assuming axial symmetry) with a rotational-like structure built upon it. In addition, the magnitude and negative sign of the $\left<2^+_2||E2||2^+_2\right>$ element is consistent with an oblate deformation of the $2^+_2$ state, which is interpreted as the $\gamma$-vibrational bandhead ($K=2$). The strong $B(E2)$ values coupling the $2^+_1$ and $4^+_2$ levels to the $2^+_2$ state support both a collective picture for the $\gamma$ band and the strong mixing between the $2^+_1$ ($K=0$) and $2^+_2$ ($K=2$) states, herewith suggesting, in addition, that the ground state and the associated rotational band are axially asymmetric. Furthermore, the $B(E2;2^+_2\to2^+_1)/B(E2;2^+_1\to0^+_1)$ ratio of $1.0(3)$, is closer to the triaxial ($\gamma = 30^\circ$) limit of $1.43$ and does not support the zero value expected in the case of axial symmetry.
    
   \subsection{Rotor Model Calculations}
    
        To investigate the role of triaxiality and provide insight into the nature of the low-lying $^{70}$Ge states, the experimental matrix elements were compared to results of calculations carried out within the framework of the generalized version of the triaxial rotor model (GTRM) \cite{Wood2004,Allmond2009,Kulp2006,Allmond2008, Allmond2010}. The latter employs independent inertia and electric-quadrupole tensors, a departure from the standard use of irrotational flow moments of inertia as used in the Davydov-Filippov (DF) model  \cite{Davydov1958}. This approach reduces systematic deviations associated with the use of fixed moments of inertia \cite{Wood2004}. Within the GTRM framework, the $E2$ matrix elements for states within the ground and $\gamma$ bands are determined analytically with three model parameters -- the quadrupole deformation $Q_0$, the asymmetry (triaxiality) of the electric quadrupole tensor $\gamma$, and the mixing angle of the inertia tensor $\Gamma$ -- using three experimental matrix elements as prescribed in Ref. \cite{Wood2004}. In the present study, these parameters were determined with the $\left<2^+_1||E2||0^+_1\right>$, $\left<2^+_2||E2||0^+_1\right>$, and $\left<2^+_2||E2||2^+_1\right>$ elements, resulting in values of $Q_0=1.35(1)$, $\gamma = 37.0\,(7)^\circ$, and $\Gamma = -44.1\,(7)^\circ$.
        
\begin{figure}[htbp!]
            \centering
            \includegraphics[width=\columnwidth]{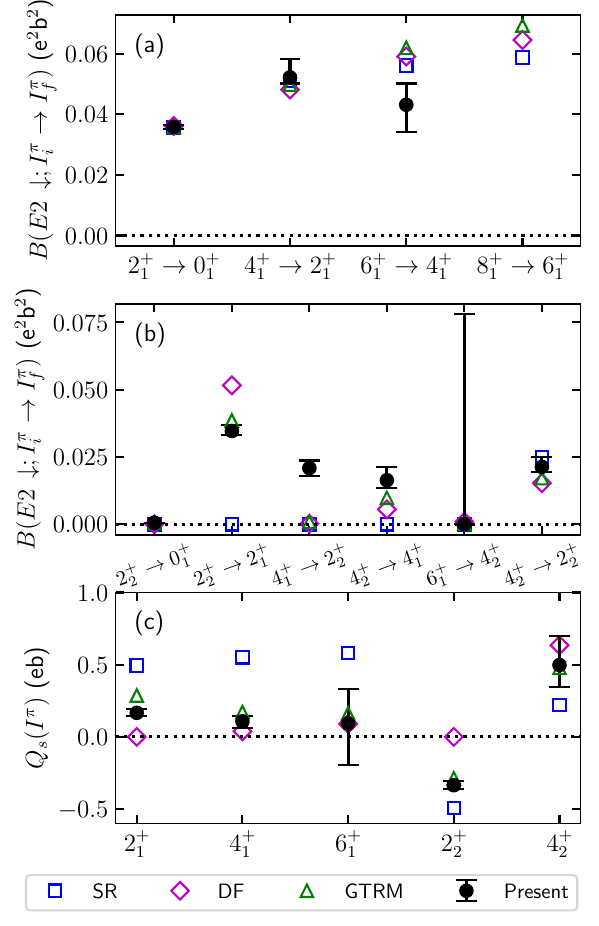} 
            \caption{(Color online) Plots of the $B(E2)$ values for selected ground-state-, $\gamma$-, and $0^+_2$-band transitions and spectroscopic quadrupole moments (diagonal $E2$ transitions) for ground-state- and  $\gamma$-band levels. The experimental data is compared to results from the symmetric rotor (SR), Davydov-Filippov (DF), and generalized triaxial rotor (GTRM) models. Note that cross-band transitions are not allowed in the SR model. Plots (a) and (b) show transitions in and between the ground-state and $\gamma$ bands, while plot (c) presents the spectroscopic quadrupole moments of states relevant to the rotor models.}
            \label{fig:BE2rotor}
        \end{figure}

        The results of the GTRM calculations are presented in Fig. \ref{fig:BE2rotor}. For comparison, results from the symmetric rotor (SR) and DF models are also plotted. All three models satisfactorily reproduce the ground-state and $\gamma$ intraband transitions and are closer to the present experimental values than to the results in Ref. \cite{Sugawara2003}, where available. This is not surprising, as all three models provide for non-zero body projections on the symmetry axis, leading to separate level sequences with $K$ values of 0 (ground-state band) and 2 ($\gamma$ band) whose transitions are predominantly governed by the degree of quadrupole deformation, $\beta$. However, the SR model predictably fails to produce interband transitions, as it assumes no mixing between the two bands. Comparatively, the GTRM and DF model more closely reproduce the $B(E2)$ values, specifically in the ground-$\gamma$ interband transitions - the only exception being the $4^+_1\to2^+_2$ transition. Of the three models, GTRM most closely predicts the spectroscopic quadrupole moments $Q_s$ in both the ground-state and $\gamma$ bands. While GTRM gives a higher $Q_s(2^+_1)$ value of $0.293$ eb compared to the measured $0.22(4)$ eb one, the moments for all other states are very closely replicated. Taken all together, this first analysis supports triaxial deformation of the $^{70}$Ge ground state, and corroborates the empirical signatures discussed in Section \ref{sec:matrixelements}.
        
    \subsection{Large-Scale Shell-Model Calculations and Rotational Invariants}\label{sec:shellmodel}
    
\begin{figure}[htbp]
            \centering
            \includegraphics[width=\columnwidth]{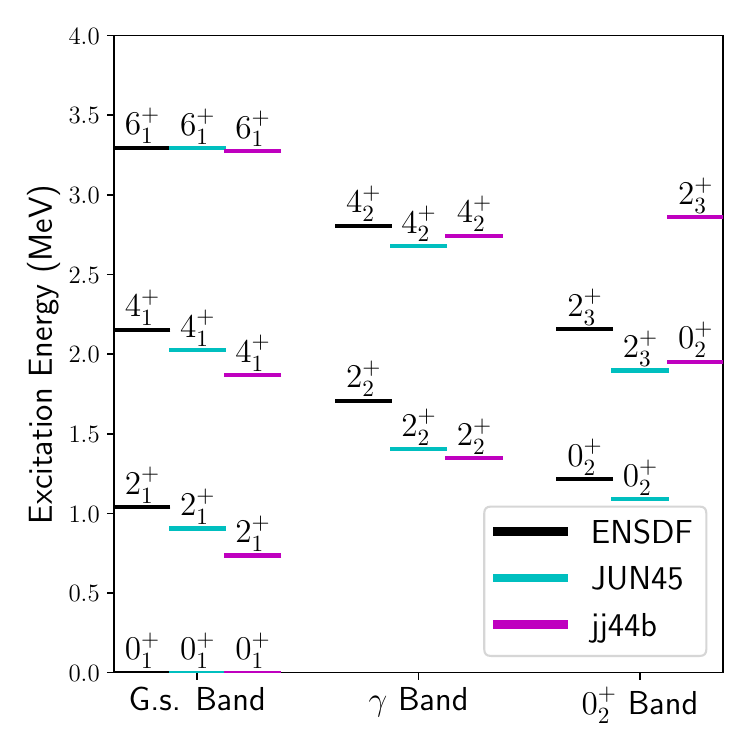} 
            \caption{(Color online) Calculated level energies for selected low-lying states in $^{70}$Ge using the JUN45 and jj44b interactions compared to ENSDF adopted values \cite{Gurdal2016}.}
            \label{fig:energiesshell}
        \end{figure}

        The experimental matrix elements and reduced transition probabilities, $B(E2)$, were also compared with results of large-scale configuration interaction calculations in the jj44 model space which comprises the $0f_{5/2}$, $1p_{3/2}$, $1p_{1/2}$, and $0g_{9/2}$ orbitals -- for both protons and neutrons -- and an inert $^{56}$Ni core. The calculations were carried out using the shell-model codes \textsc{NuShellX}~\cite{BROWN2014115} and \textsc{KShell}~\cite{kshell}. Two effective interactions, tuned for the $f_{5/2}pg_{9/2}$ model space -- the jj44b and JUN45 Hamiltonians -- were used; these have been extensively applied to study low-lying structures in nuclei within the $A \approx 60 - 70$ mass region and have successfully reproduced the general features observed in the experimental data. The computed excitation energies for the yrast and near-yrast states are compared with the experimental data in Fig.~\ref{fig:energiesshell}. The excitation energies are systematically lower than the ENSDF adopted values with rms deviations of about 170 keV for the JUN45 interaction and 410 keV for jj44b, which are within the expected deviations for such interactions. The major exceptions are the $0^+_2$ and $2^+_3$ states, for which the jj44b interaction overestimates the adopted energies by about $150\,\%$.

\begin{figure}[htbp]
            \centering
            \includegraphics[width=\columnwidth]{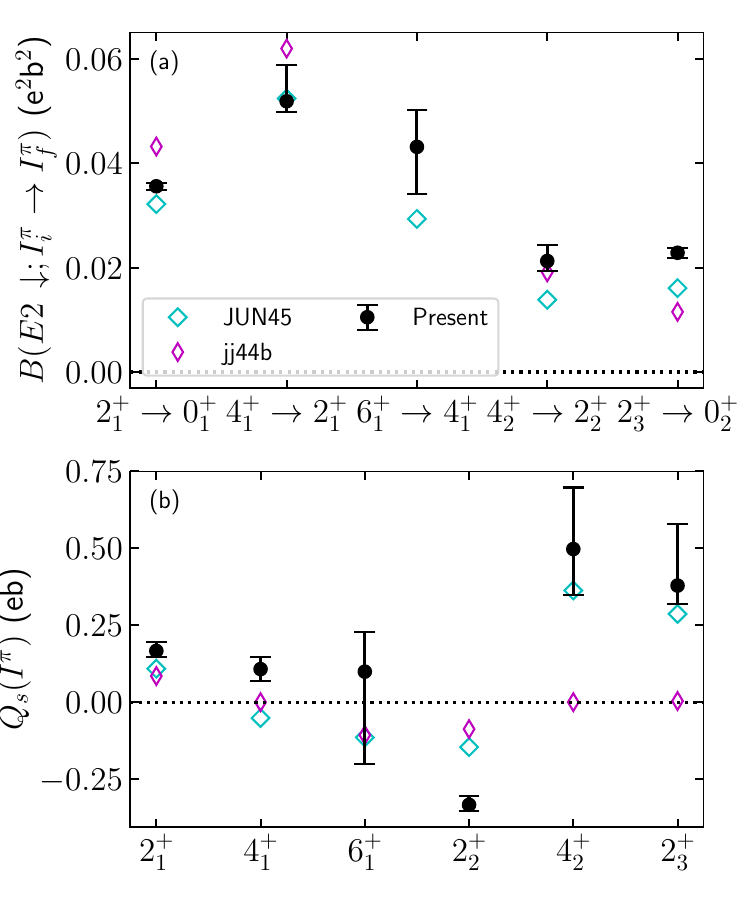} 
            \caption{(Color online) Calculated $B(E2)$ values (a) and spectroscopic quadrupole moments (b) for selected low-lying transitions and levels in $^{70}$Ge using the JUN45 and jj44b interactions compared to values derived from the present set of matrix elements. Note that the jj44b interaction did not produce a value for the $6^+_1\to4^+_1$ transition (not shown).}
            \label{fig:BE2elementsshell}
        \end{figure}

        For both interactions, isoscalar effective charges of $e_\pi = 1.8$ and $e_\nu = 0.5$ were employed. While previous studies employed various combinations of $e_\pi = 1.5, 1.8$ and $e_\nu =0.5,0.8,1.1$ in  calculating the structural properties of the even-even Ge isotopes \cite{Robinson2011,Kaneko2015,JUN45,Hirsch2012}, the selected values were used because they most closely agree with the present experimental $B(E2)$ strengths for the ground-state band. The computed reduced transition probabilities and spectroscopic quadrupole moments, $Q_s(I^\pi)$, for transitions and states important for the rotor models are presented in Figs. \ref{fig:BE2elementsshell} (a) and (b), respectively. Both interactions replicate the overall trend in the experimental  $B(E2)$ values for the ground-state- and $\gamma$-band transitions as well as for structures associated with the $0^+_2$ state. The notably large $B(E2\downarrow;2^+_3\to0^+_2)$ value reported by Ref. \cite{Sugawara2003} was observed to be substantially reduced in the present measurement, thereby bringing both interactions in better agreement with the experimental value. The situation differs somewhat for the diagonal matrix elements and, consequently, for the quadrupole moments, $Q_s(I^\pi)$. While JUN45 more closely predicts the magnitudes of the spectroscopic quadrupole moments for the low-lying states, both interactions predict a sign flip in the $Q_s$ values within the ground-state band; they predict prolate instead of oblate deformation for the $4^+_1$ and $6^+_1$ levels, in contradiction with experiment. This discrepancy in sign suggests a divergence in the predicted deformation behavior relative to the observed data. It is also should be noted that the jj44b interaction significantly underestimates the $Q_s$ values for the $4^+_{1,2}$ and $2^+_3$ states, predicting spherical charge distributions and, thus, underscoring the failure of the interaction to accurately describe the collective quadrupole properties of these levels.
        
\begin{figure}[htbp]
            \centering
            \includegraphics[width=\columnwidth]{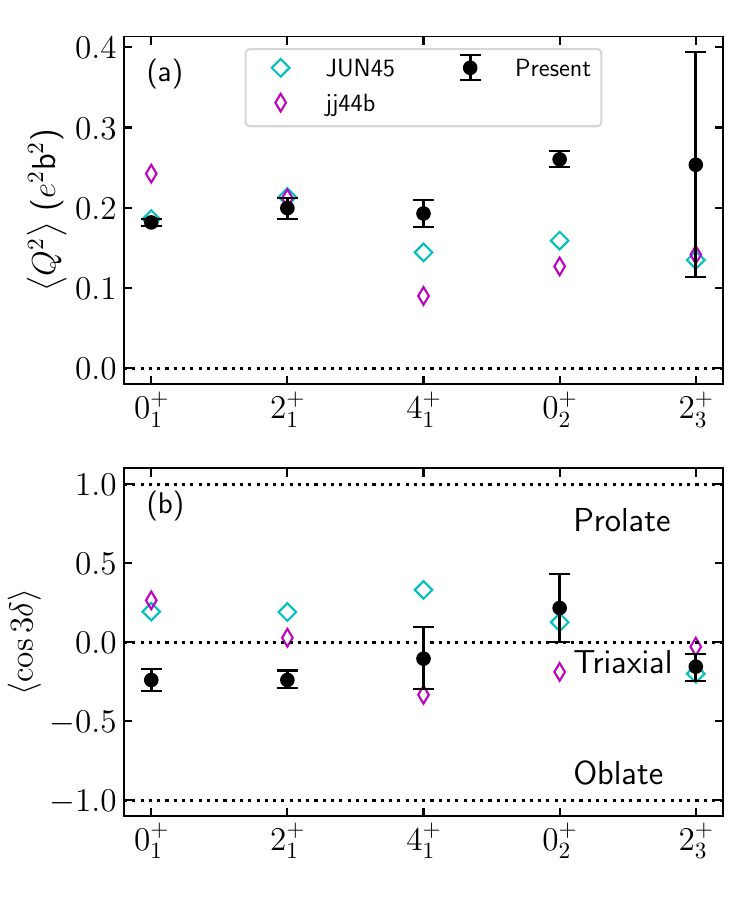}
            \caption{(Color online) Comparison of the calculated values of the $\langle Q^2\rangle$ (a) and $\langle \cos 3\delta\rangle$ (b) rotational invariants for selected low-lying states in $^{70}$Ge using the JUN45 and jj44b interactions with values from the present set of matrix elements.}
            \label{fig:invariantsshell}
        \end{figure}

        To gain further insight into the deformation characteristics of the $^{70}$Ge low-lying states, the measured $E2$ matrix elements were used to compute rotational invariant quantities according to the Kumar-Cline sum rules \cite{KK1,KK2,KK3}. The advantage of this framework is that the deduced invariant quantities are a model-independent measure of the quadrupole deformation of the nuclear charge distribution in the intrinsic frame of the nucleus. Without loss of generality, these invariants can be expressed in terms of two deformation parameters, $Q$ and $\delta$, which are analogous to the elongation parameter $\beta$ and the asymmetry angle $\gamma$ in the Bohr Hamiltonian \cite{Bohr1975}. The experimental expectation values $\langle Q^2 \rangle$ and $\langle\cos3\delta\rangle$ describe the magnitude of the average symmetric quadrupole deformation and degree of axial asymmetry of a charged ellipsoid, respectively. These are related to the Bohr variables as follows:
        \begin{eqnarray}
            \beta_{\mathrm{rms}} = \frac{4\pi}{3ZeR^2}\sqrt{\langle Q^2\rangle},\\
            \gamma = \frac{1}{3}\arccos\langle\cos3\delta\rangle,
        \end{eqnarray}
        where $R=1.2A^{1/3}$ fm, $e$ is the elementary charge, and $Z$ and $A$ are the atomic and mass numbers, respectively. The invariant quantities computed from the present matrix elements are compared with those obtained with the JUN45 and jj44b interactions in Fig. \ref{fig:invariantsshell}. To test convergence, these values were recomputed while including GTRM matrix elements (where available) for couplings which could not be determined from the fit. Changes of $5\,\%$ or less were considered to correspond to convergence. The invariants reported are those computed from the fitted matrix elements alone.
        
        The most obvious feature of the quadrupole shape invariant, $\langle Q^2\rangle$, displayed in Fig. \ref{fig:invariantsshell}, is the nearly constant behavior with spin, which is only reproducible by a collective rotor model. This is unexpected, as the $E(4^+_1)/E(2^+_1)$ value of $\approx 2$ for $^{70}$Ge is indicative of a vibrational nucleus. Note that the large $\langle Q^2\rangle$ of $\approx 0.64$ e${^2}$b${^2}$ for the $0^+_2$ level reported by Sugawara \textit{et al.,} \cite{Sugawara2003}, which was driven by the large $B(E2\downarrow; 2^+_3\to0^+_2)$ value, has been determined to be considerably smaller in the present study. This large $\langle Q^2\rangle$ value corresponds to a quadrupole deformation of $\beta_\mathrm{rms}\approx 0.43$, a degree of deformation unusually large for level sequences at low excitation energies in this mass region. Typically, a $0^+$ state of such sizable deformation would serve as the band head of a highly-deformed or superdeformed rotational band. However, evidence of such a band has not been observed to date. Furthermore, empirical signatures such as the energy of the $2^+_3$ level and the intraband $B(E2)$ values, suggest that the $0^+_{1,2}$ states have a similar degree of deformation. This picture is corroborated by both interactions, which predict average $\langle Q^2\rangle$ values of $\approx 0.19$ e$^{2}$b$^{2}$ ($\beta_\mathrm{rms}\approx 0.23$) and $\approx 0.26$ e$^{2}$b$^{2}$  ($\beta_\mathrm{rms}\approx 0.27$) for the $0^+_{1,2}$ levels, respectively.

\begin{figure}[!b]
            \centering
            \includegraphics[width=\columnwidth]{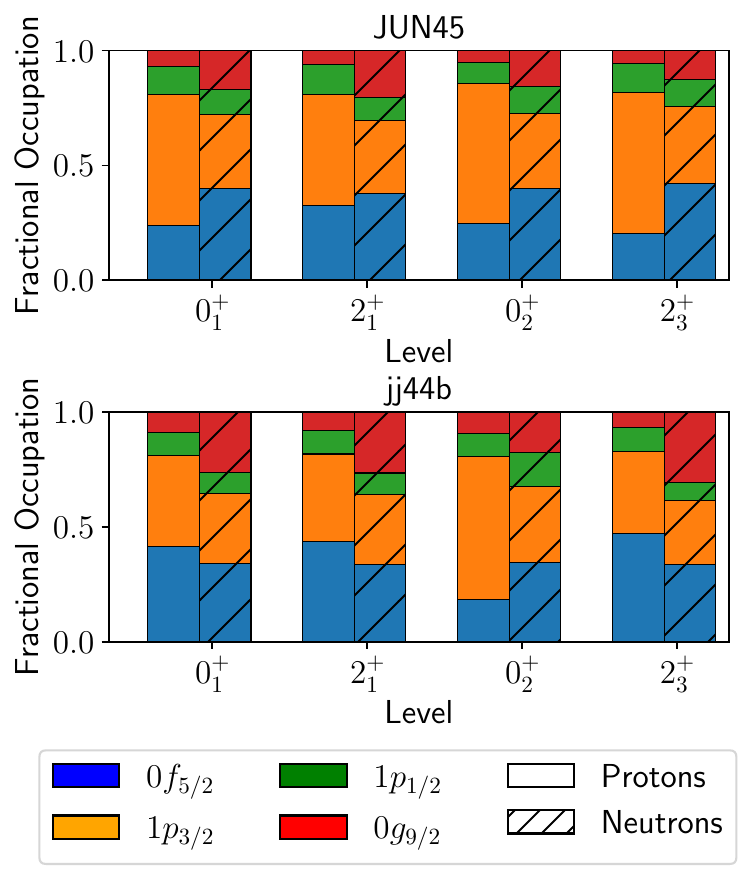} 
            \caption{(Color online) Calculated occupation numbers for selected low-lying states in $^{70}$Ge using the JUN45 and jj44b interactions. The occupation numbers are presented as fractions. The similarity in the occupation numbers between the two $0^+$ states suggests that they are of similar character, herewith corroborating the results of the mixing calculations. Note the significant population of the $0g_{9/2}$ neutron orbital, which drives quadrupole deformation in this region.}
            \label{fig:occnums}
        \end{figure}

        A more complex picture emerges when comparing the experimental and theoretical $\langle\cos3\delta\rangle$ values. In the ground state band, the data indicate a small, negative value which remains roughly constant with increasing spin. The average $\langle\cos3\delta\rangle$ value is approximately $-0.19$ for the ground-state band and $0.03$ for the $0^+_2$ sequence. These correspond to $\gamma_\mathrm{rms}$ values of $34^\circ$ and $29^\circ$, respectively. These values indicate a near-maximum triaxial deformation at low energy, further reinforcing the essential role of triaxiality in determining the structure of even-even Ge isotopes, in agreement with conclusions reached in Refs. \cite{Ayangeakaa2016,nirupama2024,Ayangeakaa2019,Ayangeakaa2023}. While both the JUN45 and jj44b predictions generally agree on the magnitude of the $\langle\cos3\delta\rangle$ parameter across the states and maintain a predominantly triaxial configuration, neither model correctly predicts all the signs of the $\expval{\cos3\delta}$ values which were deemed to have converged. The JUN45 interaction computes positive $\langle\cos3\delta\rangle$ values (implying an oblate-triaxial deformation) for nearly all levels in the ground-state and $0^+_2$ bands, with the exception of the $2^+_3$ state. In contrast, the jj44b Hamiltonian suggests a negative sign (prolate-triaxial deformation) for most states, indicating a notable difference in the underlying deformation tendencies predicted by each interaction. The similarity in the $\beta_\mathrm{rms}$ and $\gamma$ values for the $0^+_{1,2}$ states is in marked contrast to the results of Ref. \cite{Sugawara2003} and suggests a less pronounced overall picture for shape coexistence in $^{70}$Ge than previously suggested. To explore the underlying structure of these levels further, the occupation numbers of protons and neutrons in the $0f_{5/2}$, $1p_{3/2}$, $1p_{1/2}$, and $0g_{9/2}$ orbitals are presented in Fig. \ref{fig:occnums} for both interactions. The JUN45 interaction predicts a nearly identical fractional occupation for both protons and neutrons in the $0^+_{1,2}$ and $2^+_{1,3}$ states. Although the jj44 Hamiltonian computes similar fractions, it highlights a subtle difference in the nucleon occupancy for the $0^+$ states: the proton $0f_{5/2}$ orbital in the $0^+_1$ state is replaced by the $1p_{3/2}$ one in the $0^+_2$ level. In addition, calculations with the JUN45 interaction reveal the same dominant $\pi(p_{3/2})^4\otimes\nu(p_{3/2}^4f_{5/2}^4p_{1/2}^2)$ configuration for the two $0^+$ states. The wave functions are, however, highly mixed, with the major components having amplitudes of 8.3\,\% and 16.5\,\%  for  the $0^+_1$ and  $0^+_2$ states, respectively. This mixing becomes even more pronounced with spin, as the leading components increasingly contribute less than 5\,\% to the total wave function. These results, combined with the insight gained from the invariant shape analysis and the empirical indicators such as the associated $2^+$ level energies, support the interpretation of the $0^+_1$ and $0^+_2$ states being rather similar in character. This structural similarity suggests the presence of strong mixing, leading to essentially identical charge distributions and similar shapes, further challenging conclusions of a distinct shape coexistence between a moderately oblate deformed $0^+_1$ ground state and a highly-deformed excited $0^+_2$ state.

    \subsection{Shape Coexistence and Two-State Mixing}
    
        To further investigate the nature of the $0^+_{1,2}$ states and explore the role of configuration mixing further, a two-state mixing calculation between the $0^+_{1,2}$ and $2^+_{1,3}$ states was performed using the formalism described in Ref. \cite{Clement2007}. In this framework, the observed $0^+_{1,2}$ and $2^+_{1,3}$ levels are interpreted as resulting from mixing between two competing configurations -- a quasi-spherical and a deformed ``intruder'' one, the latter arising from the neutron occupation of the $0g_{9/2}$ orbital. Consequently, the experimentally observed $0^+_{1,2}$ and $2^+_{1,3}$ levels are expressed as linear combinations of the pure (unperturbed) states corresponding to each configuration:
        \begin{eqnarray*}
            |0^+_1\rangle &=& \cos\theta_0|0^+_{\text{sph}}\rangle + \sin\theta_0|0^+_{\text{int}}\rangle, \\
            |0^+_2\rangle &=& -\sin\theta_0|0^+_{\text{sph}}\rangle + \cos\theta_0|0^+_{\text{int}}\rangle, \\
            |2^+_1\rangle &=& \cos\theta_2|2^+_{\text{sph}}\rangle + \sin\theta_2|2^+_{\text{int}}\rangle, \\
            |2^+_3\rangle &=& -\sin\theta_2|2^+_{\text{sph}}\rangle + \cos\theta_2|2^+_{\text{int}}\rangle,
        \end{eqnarray*}
        where $|0^+_{\text{sph}}\rangle$ and $|2^+_{\text{sph}}\rangle$ represent the unperturbed states of the quasi-spherical configuration, while $|0^+_{\text{int}}\rangle$ and $|2^+_{\text{int}}\rangle$ denote those associated with the unperturbed intruder one. The parameters $\theta_{0,2}$ are the mixing angles for the $0^+$ and $2^+$ states, respectively. 
        
        In the intrinsic unmixed frame, $E2$ transitions between pure states of different configurations are forbidden. That is, $\langle I'_{\text{sph}}||E2||I_{\text{int}}\rangle = 0$
        for all $I$, $I'$ values. The reduced matrix elements for the observed transitions can be written in terms of the mixing angles and the intrinsic matrix elements, resulting in a system of four equations (see Ref. \cite{Clement2007}). From these, two sets of solutions  involving the mixing amplitudes, the unperturbed energy separation of the $0^+$ states, and the interaction matrix element can be calculated. The results, labeled positive and negative, are presented in Table \ref{tab:mixingcalc}, along with those calculated using the JUN45 \cite{JUN45} and jj44b \cite{jj44b} interactions.
         \begin{table}[!t]
            \caption{Results of a two-state mixing calculation between the $0^+_{1,2}$ and $2^+_{1,3}$ states of $^{70}$Ge. The mixing angle of the two $0^+$ states, $\theta_0$, and the two $2^+$ states, $\theta_2$, are calculated from the present set of matrix elements and the results of the jj44b and JUN45 calculations. The interaction matrix element, $V_0$ and the energy separation of the pure $0^+$ states $\Delta E_0^{\text{pure}}$ are given in units of keV, while the predicted (denoted with a hat) spectroscopic quadrupole moments and $E2$ matrix elements are given in units of eb.}
            \label{tab:mixingcalc}
            \begin{ruledtabular}
            \begin{tabular}{cccccccccc}
                \multicolumn{1}{c}{\multirow{2}{*}{}} & \multicolumn{2}{c}{Present} && \multicolumn{2}{c}{JUN45} && \multicolumn{2}{c}{jj44b} \\
                \cline{2-3}\cline{5-6}\cline{8-9}
                \multicolumn{1}{c}{} & \multicolumn{1}{c}{$+$} & \multicolumn{1}{c}{$-$} && \multicolumn{1}{c}{$+$} & \multicolumn{1}{c}{$-$} && \multicolumn{1}{c}{$+$} & \multicolumn{1}{c}{$-$}  \\
                \hline
                $\cos^2\theta_0$ & $0.49$ & $0.51$ && 0.68 & 0.32 && 0.97 & 0.03 \\
                \rule{0pt}{3ex}
                $\cos^2\theta_2$ & $0.20$ & $0.80$ && 0.91 & 0.09 && 1.0 & 0.0 \\
                \rule{0pt}{3ex}
                $|V_0|$ & \multicolumn{2}{c}{608} && \multicolumn{2}{c}{568} && \multicolumn{2}{c}{223} \\
                \rule{0pt}{3ex}
                $|\Delta E_0^\text{pure}|$ & \multicolumn{2}{c}{20.3} && \multicolumn{2}{c}{415.4} && \multicolumn{2}{c}{1131.9} \\
                \rule{0pt}{3ex}
                $\hat{Q}_s(2^+_1)$ & 0.325 & -0.325 && -0.353 & 0.353 && -0.428 & 0.428 \\
                \rule{0pt}{3ex}
                $\hat{Q}_s(2^+_3)$ & -0.124 & 0.124 && 0.179 & -0.179 && 0.219 & -0.219 \\
                \rule{0pt}{3ex}
                $\big<2^+_1\big|\big|\hat{E2}\big|\big|2^+_3\big>$ & -0.385 & 0.385 && -0.247 & 0.247 && 0.017 & -0.017 \\
            \end{tabular}
          \end{ruledtabular}
        \end{table}
        
        Both solutions derived from the data suggest near-maximum mixing of the two $0^+$ states, as indicated by the squared mixing amplitude of $\cos^2\theta_0=0.51$ for the negative solution ($\cos^2\theta_0=0.49$ for the positive one). The negative solution, however, suggests a less deformed configuration for the $2^+_1$ state and a predominantly deformed one for the $2^+_3$ level, in agreement with the experimental values of the diagonal $E2$ matrix elements. The positive solution implies the reverse. Hence, the negative solution is adopted, as it better reproduces the experimental data. The significant mixing of the $0^+$ states is most closely replicated by the JUN45 calculations. In addition, both the experimental data and the calculations indicate that the purity of the wavefunctions increases rapidly with spin, as evidenced by the $\cos^2\theta$ values being near unity for the $2^+$ states. By predicting a strong degree of mixing, the JUN45 results support the notion that the nearly identical $0^+_{1,2}$ occupation numbers are likely a result of mixing between the intrinsic configurations, resulting in similar structural properties. 
        
        In the context of the two-state mixing model, the quadrupole moments for the $2^+_1$ and $2^+_3$ states are predicted to be opposite in shape, which is observed experimentally. However, the calculation also finds the $2^+_1$ state to be more deformed than the $2^+_3$ one, which is not supported by the data. It should be mentioned that, given the evidence of near-maximum triaxiality seen in both the $0^+_1$- and $0^+_2$-state configurations, a more complete treatment of the mixing between the two would include each configuration's corresponding $\gamma$ ($K=2$) bands, which could reduce the deviation from the experimental results. Therefore, these results should be taken as a first-order approximation of the properties of the mixed states and the couplings between them.
        
    \subsection{Relativistic Density Functional Theory and the Five Dimensional Collective Hamiltonian}\label{sec:rdft-gbh}
    
        The collective quadrupole properties of the low-lying states in $^{70}$Ge were further analyzed using constrained relativistic density functional theory (RDFT)~\cite{J.Meng2006PRC, J.Meng2006PPNP, J.Meng2011PIP, J.Meng2016book}, which describes a wide array of nuclear properties, including ground-state deformations and collective excitations of medium and heavy nuclei. RDFT is formulated in the context of relativistic mean-field theory, where nucleons are described as Dirac particles which interact through effective meson fields. The calculations were carried out using the PC-PK1 effective interaction~\cite{P.W.Zhao2010PRC}, with the effects of pairing correlations neglected for simplicity. The potential energy surface (PES) was obtained by solving the Dirac equation using a three-dimensional harmonic oscillator basis with 12 major oscillator shells. In addition, a density-independent $\delta$ force was included in the particle-particle channel, with strength parameters of $349.5~\textrm{MeV}\textrm{fm}^3$ for neutrons and $330.0~\textrm{MeV}\textrm{fm}^3$ for protons and the $(\beta,\gamma)$ deformation parameters were varied within the specified intervals: $0.0 \leq \beta \leq 0.6$ and $0^\circ \leq \gamma \leq 60^\circ$, in step sizes of $\Delta\beta=0.05$ and $\Delta\gamma=6^\circ$, respectively. 
        
\begin{figure}[t]
        	\centering
        	\includegraphics[width=\columnwidth]{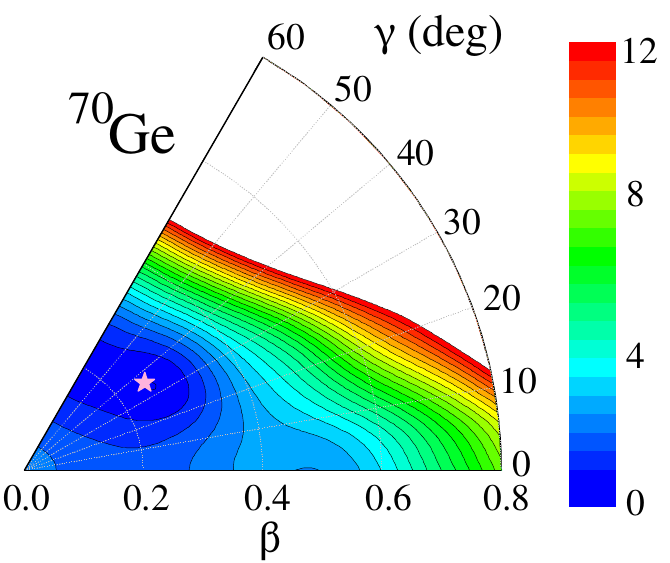}
        	\caption{(Color online) Potential energy surface in the $\beta$-$\gamma$ plane ($0\leq \beta \leq 0.6$, $0^\circ \leq \gamma \leq 60^\circ$) for the ground-state configuration of $^{70}$Ge from RDFT calculations with the PC-PK1 effective interaction. Energies depicted on the plot are normalized relative to the absolute minimum energy (in MeV), denoted by a pink star symbol. Contour lines on the graph are separated by 0.5 MeV. The PES indicates that the ground state has a minimum at $\beta=0.25$, $\gamma=36^\circ$.}
        	\label{fg:PES}
        \end{figure}

        The resulting PES in the $\beta$-$\gamma$ plane for the ground-state configuration is presented in Fig.~\ref{fg:PES}, where the energy values are normalized to the binding energy of the absolute minimum. The latter minimum corresponds to the ground state and is represented by a pink star in the figure. It indicates that the ground-state configuration of $^{70}$Ge favors a moderate oblate deformation with $\beta=0.25$ and a soft, but significant, degree of triaxiality of $\gamma \approx 36^\circ$. This is consistent with the data as well as with the results of the analyses presented in the preceding sections.
        
\begin{figure}[t]
            \centering
            \includegraphics[width=\columnwidth]{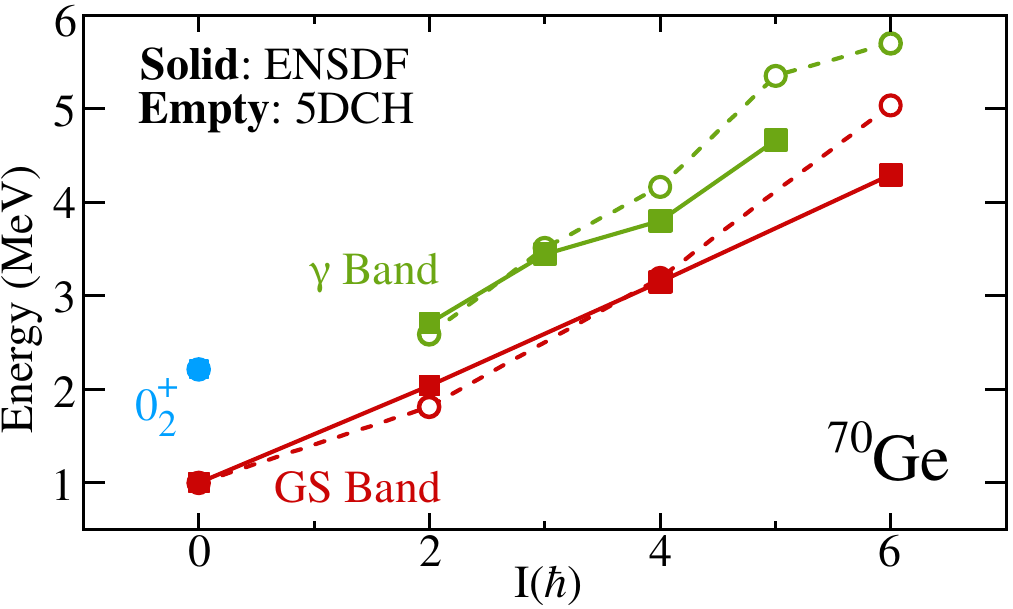}
            \caption{(Color online) Energy as a function of spin calculated using the 5DCH formalism. In these calculations, the mass parameter along the $\beta$ direction was enhanced to reproduce the energy of the $0_2^+$ state (see text).}
            \label{fig:rdft-energy}
        \end{figure}

        To investigate the collective dynamics of $^{70}$Ge, the 
        five-dimensional collective Hamiltonian (5DCH)~\cite{Niksic2009PRC, 
        Z.P.Li2009PRC_v1, Niksic2011PPNP, Z.P.Li2016JPG, Z.P.Li2023book} was employed. The collective potential as well as the inertia parameters were extracted directly from the RDFT calculations. This approach ensures a fully microscopic description of the nuclear shape dynamics, restores rotational symmetry, and accounts for triaxiality. The 5DCH equations were solved numerically by diagonalizing the collective Hamiltonian in a truncated basis of the collective model space to determine the collective energies and wave functions for each spin. To achieve accurate agreement with experimental data and, in particular, reproduce the experimental energy of the $0^+_2$ state, the mass parameter along the $\beta$ direction was enhanced to compensate for the inherent overestimation of the stiffness in the quadrupole deformation $\beta$ generally observed in RDFT calculations. As seen in Fig.~\ref{fig:rdft-energy}, the experimental energy spectra from the ground state up to $I=6\hbar$ are relatively well reproduced by the 5DCH calculations. In fact, while the energy of $0^+_2$ is almost exactly calculated, slight deviations are observed at higher angular momentum. Overall, these results confirm the collective nature of the low-spin structure. 
        
        Furthermore, the collective wave functions determined from the 5DCH formalism were subsequently used to compute the rotational invariant, $\langle Q^2 \rangle$ and $\langle \cos3\delta \rangle$, values following the method described in Ref.~\cite{Srebrny2006NPA}. These are presented in Table~\ref{tab:rdft-ghb} for the $0^+_1$ ground state and excited $0^+_2$ level, along with the corresponding quadrupole deformation parameters. The sets of 5DCH computed deformation parameters agree reasonably well with the experimental ones. In particular, the calculated $\langle Q^2 \rangle$ invariant agrees with the experimental value of $0.260(1)$ e$^2$b$^2$ for the $0^+_2$ state. Likewise, the $\langle Q^2 \rangle$ value for the ground state is reproduced within uncertainty. The latter is, however, computed with a larger fluctuation, indicating a significant degree of softness. In the same fashion, the central values of $\langle \cos3\delta \rangle$, calculated within the 5DCH framework, are in good agreement with the experimental values for both states. The correspondence between calculations and data reinforces the notion of a less pronounced, but distinguishable, coexistence between a predominantly oblate-triaxial ground state and a prolate-leaning excited $0^+_2$ state with significant triaxiality.
        
        \begin{table}[]
        \centering
        \caption{Comparison of the average rotational invariants deduced from the experimental matrix elements with those calculated using the five dimensional collective Hamiltonian (5DCH) formalism for the $0_1^+$ and $0_2^+$ states. The $\langle Q^2\rangle$ values are given in units of e$^2$b$^2$. The theoretical uncertainties represent a $1\sigma$ fluctuation in the calculated values. It should be noted that the $0^+_2$ state has two maxima in the probability density function (see Fig. \ref{fg:Prob}); however, only the global (prolate) maximum is listed here.}
        \label{tab:rdft-ghb}
            \begin{ruledtabular}
                \begin{tabular}{cccccc}
                    \multirow{2}{*}{} & \multicolumn{2}{c}{$0_1^+$} && \multicolumn{2}{c}{$0_2^+$} \\ \cline{2-3} \cline{5-6} & \multicolumn{1}{c}{EXP} & 5DCH && \multicolumn{1}{c}{EXP} & 5DCH \\
                    \hline
                    $\expval{\beta}$ & 0.228(3) &0.27(5)&&  0.273(1)   &  0.35(10) \\
                    $\expval{\gamma}$ & $34.6(14)^\circ$ &$31(11)^\circ$&& $25.8(43)^\circ$ &  $15(11)^\circ$ \\
                    $\left<Q^2\right>$ & \multicolumn{1}{c}{0.182(5)} & 0.27(11) && \multicolumn{1}{c}{0.260(1)} & 0.43(21) \\ 
                    $\left<\cos 3\delta\right>$ & \multicolumn{1}{c}{-0.24(7)} & -0.03(59) && \multicolumn{1}{c}{0.22(22)} & 0.71(61) 
                \end{tabular}
            \end{ruledtabular}
        \end{table}
        
        \begin{figure}[ht]
        	\centering
        	\includegraphics[width=\columnwidth]{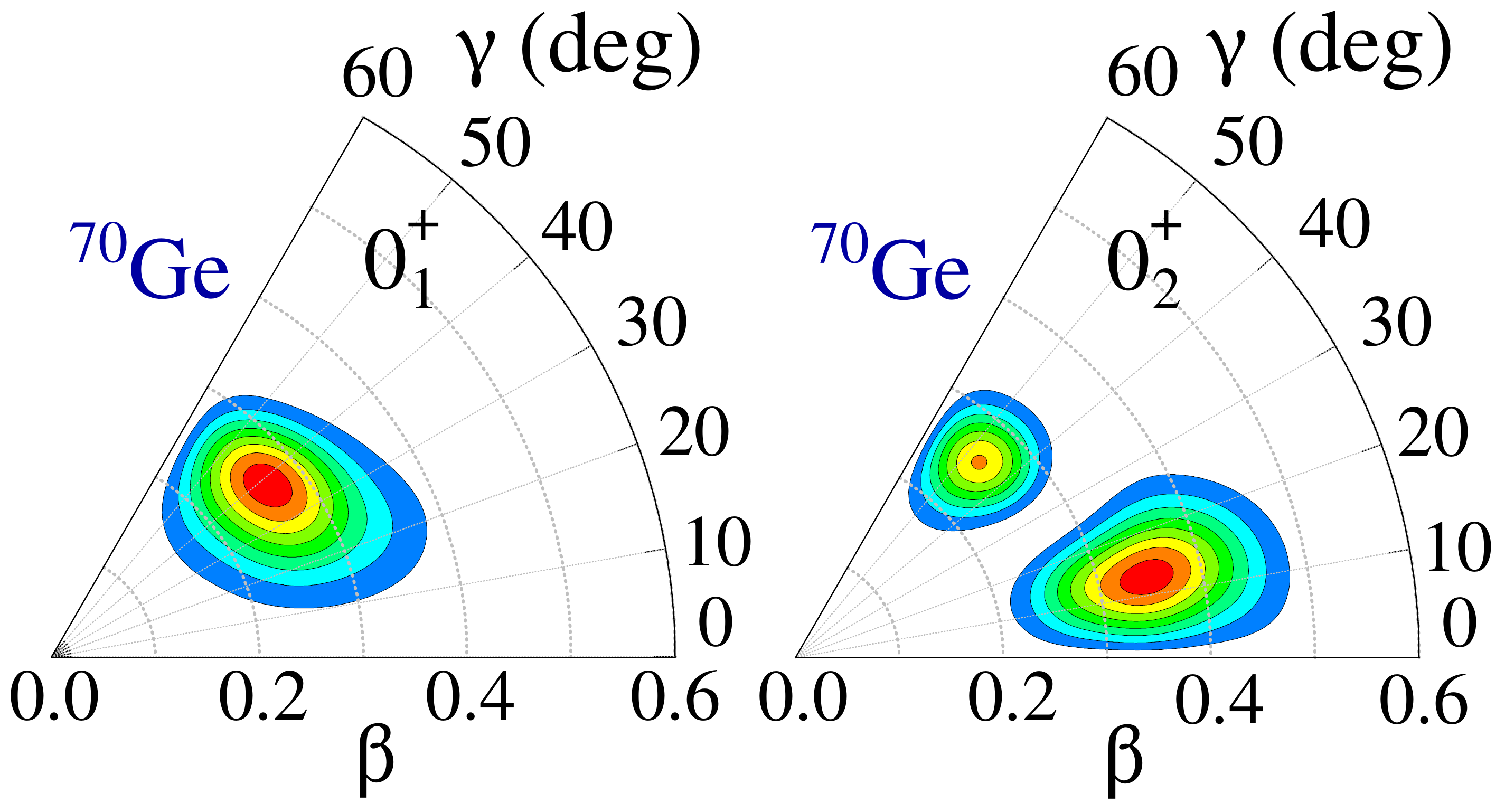}
        	\caption{(Color online) The probability density distributions of the collective wave functions in the $\beta$-$\gamma$ plane for the  $0_1^+$ and $0_2^+$ states of $^{70}\textrm{Ge}$ as calculated by the 5DCH. The maximum of the probability corresponds to the minimum of the PES for the $0_1^+$ state, while the $0_2^+$ state has a bimodal distribution with one maximum near $(\beta\approx0.35,\, \gamma\approx 15^\circ)$ and another near $(\beta \approx0.23,\, \gamma\approx 45^\circ)$. See text for a further discussion.}
        	\label{fg:Prob}
        \end{figure}
        
        The structure and dynamics of the quadrupole deformation associated with each state can be further elucidated through the probability density distribution of the collective wave functions computed using the 5DCH formalism. Figs.~\ref{fg:Prob} (a) and \ref{fg:Prob} (b) present the density distributions in the $\beta$-$\gamma$ plane for the $0_1^+$ and $0_2^+$ states of $^{70}\textrm{Ge}$ obtained in this approach. The wave function distribution for the $0_1^+$ state [Fig.\ref{fg:Prob}(a)] is consistent with the characteristics of the PES, i.e., the maximum of the probability distribution, located at ($\beta\approx0.22$, $\gamma \approx 39^\circ$), corresponds to the minimum of the RDFT-derived PES of Fig.~\ref{fg:PES}, corresponding to a well-localized and stable equilibrium triaxial-oblate configuration. 
        In contrast, the collective wave function for the $0_2^+$ state, shown in Fig.\ref{fg:Prob} (b), 
        exhibits a nodal structure with two distinct maxima in the collective probability density distribution located at ($\beta \approx 0.35$, $\gamma \approx 15^\circ$) and ($\beta \approx 0.23$, $\gamma \approx 45^\circ$). It is important to note that these probability density maxima do not correspond to separate minima in the PES, but rather reflect a potential that is notably soft along the $\beta$ degree of freedom. 
        This indicates that the $0_2^+$ state exhibits a soft extended topology in the $\beta$ direction, suggestive of a shallow potential that supports large-amplitude collective fluctuations. The more pronounced maximum at ($\beta \approx 0.35$, $\gamma \approx 15^\circ$) corresponds to a moderately deformed prolate-triaxial configuration that dominates the structure, while the second peak reflects the nodal behavior of a vibrationally excited collective mode.
        This behavior is captured by the 5DCH Hamiltonian, which accounts for large-amplitude motion in both the $\beta$ and $\gamma$ degrees of freedom, thus providing a comprehensive description of shape dynamics in $^{70}$Ge. Therefore, the probability density distributions of the collective wave functions for the $0^+_{1,2}$ states present a subtle case of shape coexistence whereby the $0^+_{1}$ state is localized within an oblate-triaxial maximum, while the $0^+_{2}$ level favors a prolate shape with significant triaxiality, consistent with the present experimental data. Finally, the calculated $E0$ transition strength, $\rho^2(E0, 0_2^+ \to 0_1^+) \times 10^3 \approx 29$, reinforces the interpretation of shape coexistence between the $0_2^+$ and $0_1^+$ states. 

\section{Conclusions} \label{sec:conclusion} 

The collective quadrupole properties of low-lying states in $^{70}$Ge were investigated by Coulomb excitation of a $^{70}$Ge beam incident on a $^{208}$Pb target. Using the state-of-the-art $\gamma$-ray tracking array, GRETINA, and the CHICO2 particle detector, $\gamma$ rays of interest were measured in kinematic coincidence with the scattered $^{70}$Ge beam. A comprehensive set of transition and static electromagnetic matrix elements was deduced from the $\gamma$-ray decay yields using the semi-classical coupled-channel code GOSIA. The resulting matrix elements expand upon the literature values and allow reduced transition probabilities, spectroscopic quadrupole moments, and rotational invariant shape parameters to be computed with increased precision. These new values were compared with the results of calculations carried out within the frameworks of the generalized triaxial rotor model, the configuration interaction shell model, and relativistic density functional theory. In general, there is good agreement between the experimental data and the theoretical calculations. In particular, the overall deformations of the $0^+_{1}$ and $0^+_{2}$ states were found to be relatively similar, but with different degrees of triaxiality. These results support the interpretation of the low-spin structure of $^{70}$Ge as a case of shape coexistence characterized by a moderately-deformed, oblate-triaxial $0^+_1$ ground state and a prolate-triaxial $0^+_2$ state. However, the coexisting configurations that comprise the $0^+_{1,2}$ states are measured to be much more similar than previously reported. This is in agreement with the general trend observed along the Ge isotopic chain, where shape coexistence, configuration mixing, and triaxiality are dominant features.
 
\section*{Acknowledgments}

This work was supported in part by the U.S. DOE, Office of Science, Office of Nuclear Physics, under Grant Nos. DE-SC0023010 (UNC), DE-FG02-97ER41041 (UNC), DE-FG02-97ER41033 (TUNL), DE-AC02-06CH11357 (ANL), DE-AC02-05CH11231 (LBNL), DE-AC52-07NA27344 (LLNL), DE-FG02-94ER40834, DE-FG02-08ER41556, DE-FG02-94ER40848, and DE-SC0020451; the NSF under Contract Nos. PHY-0606007, PHY-2011890, PHY-2208137, and PHY-2310059; the UNC Startup Funds of A. D. Ayangeakaa; the National Natural Science Foundation of China under Grant No. 12205103; and the National Key R\&D Program of China No.~2024YFE0109800 and 2024YFE0109803. Work at Los Alamos National Laboratory, operated by Triad National Security, LLC, for the National Nuclear Security Administration of the U.S. Department of Energy was done under Contract No. 89233218CNA000001. GRETINA was funded by the U.S. DOE, Office of Science, Office of Nuclear Physics, and operated by the ANL and LBNL contract numbers above. This research used the resources of Argonne National Laboratory's ATLAS facility, a DOE Office of Science User Facility.

\bibliographystyle{apsrev4-2}
\bibliography{70Ge-PRC.bib}

\end{document}